\documentclass{aa}

\usepackage{graphicx}
\usepackage[varg]{txfonts}
\usepackage{natbib}
\usepackage[pdftex]{hyperref}

\def\ImUnit{\mathbf{i}}

\def\sun{\odot}
\def\earth{\oplus}

\def\expo#1{\mathbf{e}^{#1}}

\def\dpart#1#2{\frac{\partial #1}{\partial #2}}
\def\scaled#1{\hat{#1}}

\def\modif#1{#1}

\newcommand\figI{
  \begin{figure}
    \centering
    \includegraphics[width=9cm,trim = 0cm 2cm 0cm 0cm, clip]{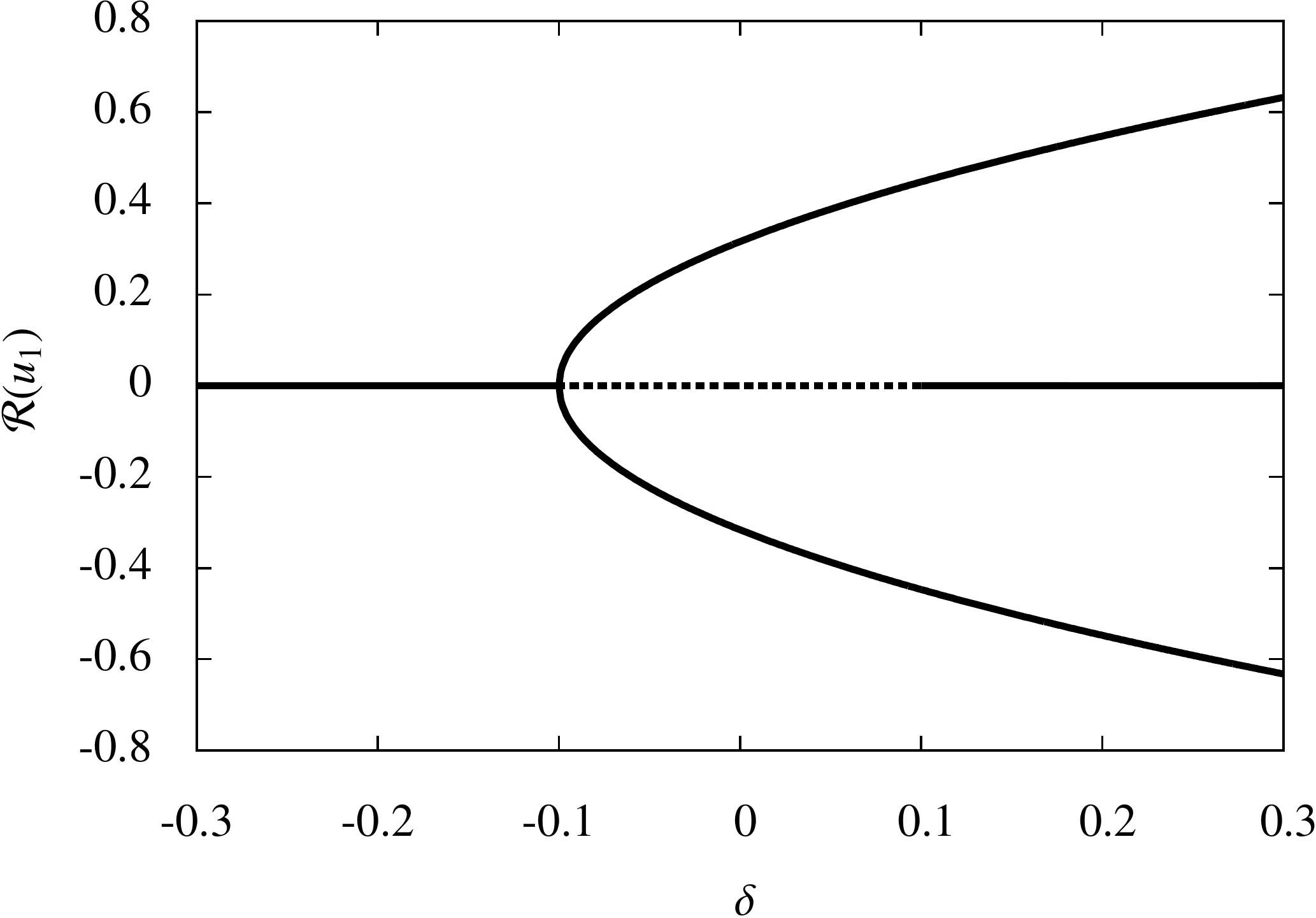}
    \includegraphics[width=9cm]{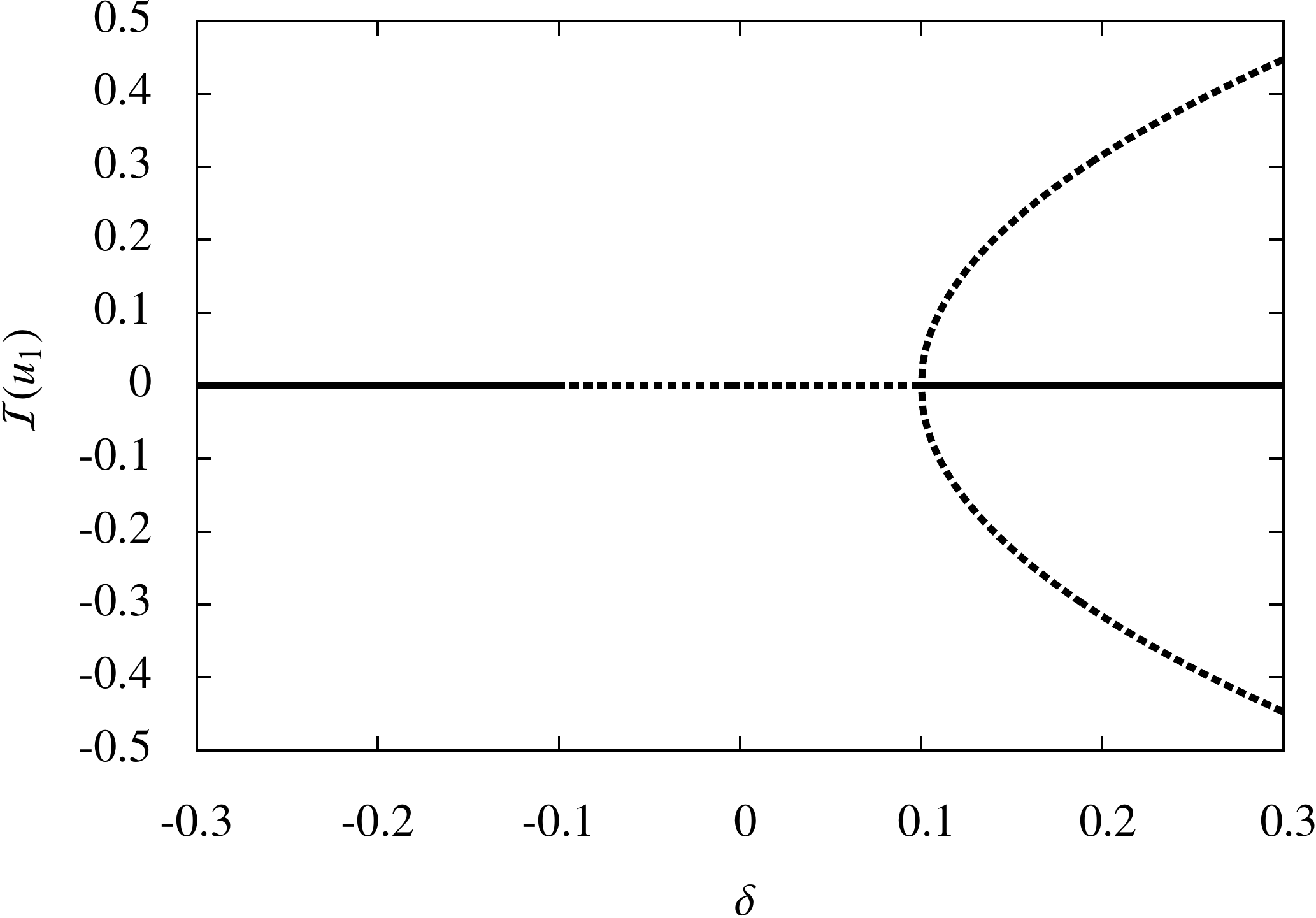}
    \caption{Position of the fixed point of a second order resonance ($q=2$)
      in the simplified model (Hamiltonian~(\ref{eq:Hqd})) as a function of $\delta$.
      We give the real (top) and imaginary (bottom) parts of $u_1$ as given by Eqs.~(\ref{eq:pfu1})
      with $R=0.1$.
      Continuous lines correspond to stable branches while dashed lines to unstable ones.}
    \label{fig:I}
  \end{figure}
}

\newcommand\figII{
  \begin{figure}
    \centering
    \includegraphics[width=8.5cm,trim = 0cm 3cm 0cm 0cm, clip]{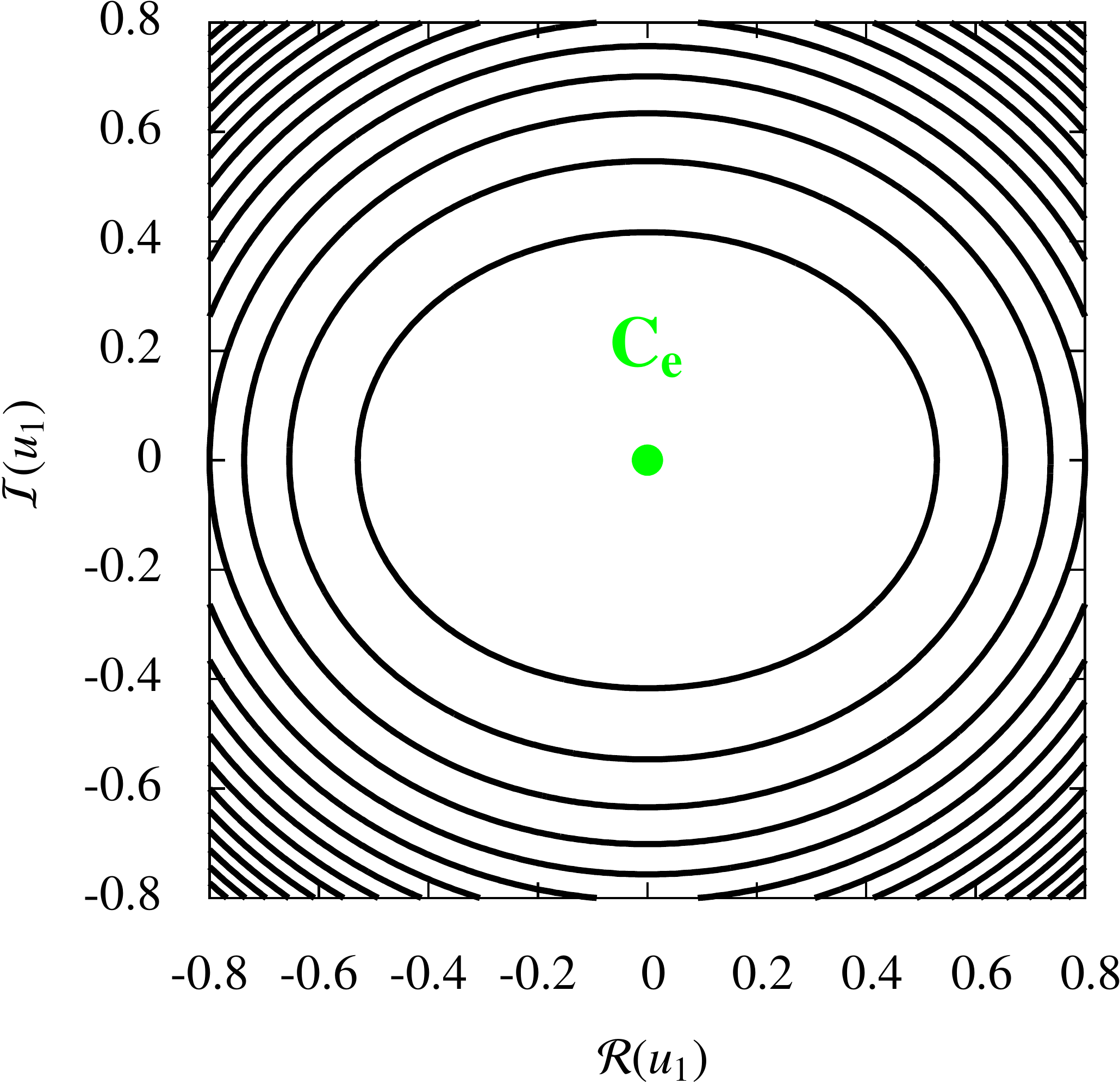}
    \includegraphics[width=8.5cm,trim = 0cm 3cm 0cm 0cm, clip]{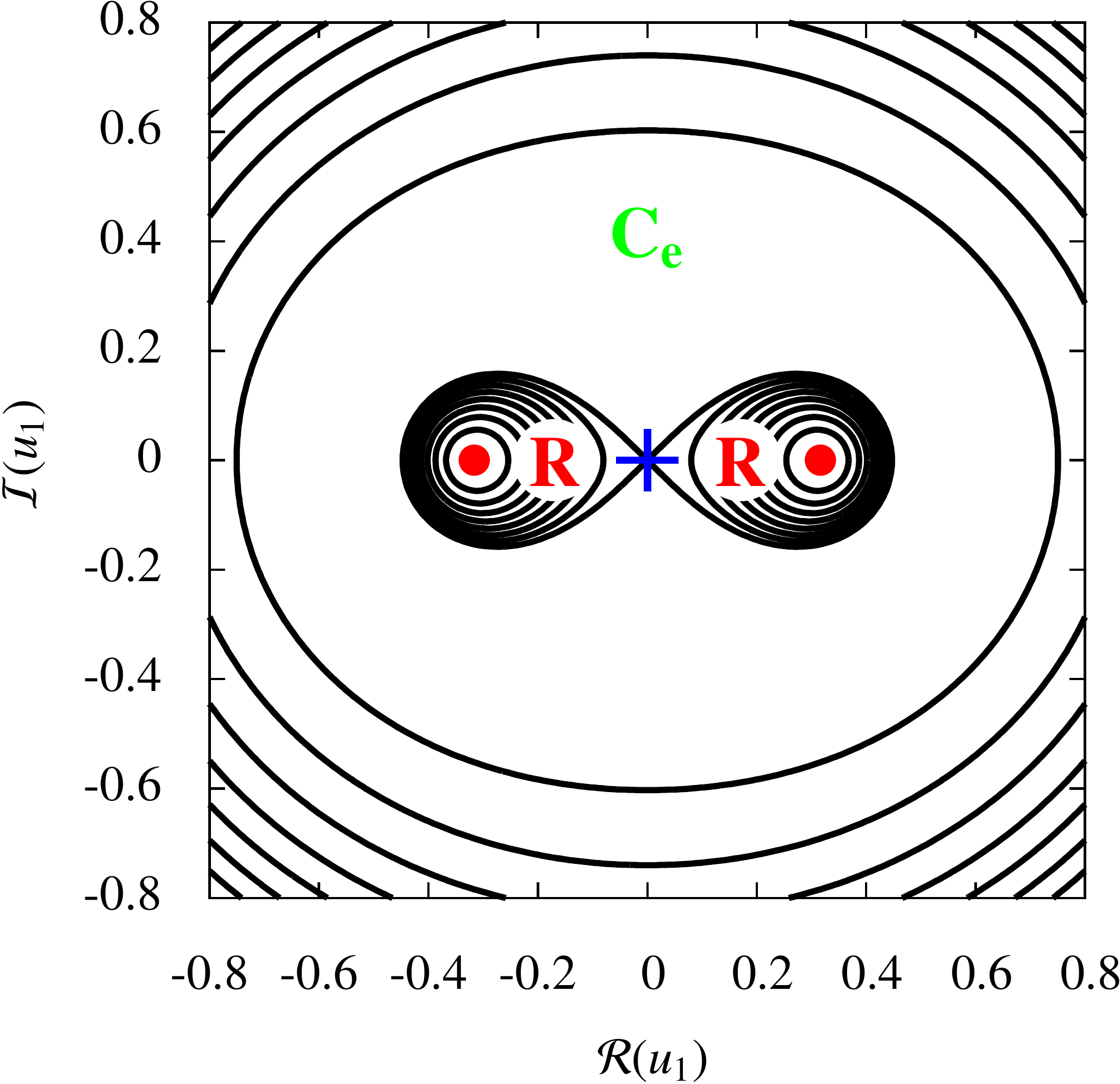}
    \includegraphics[width=8.5cm]{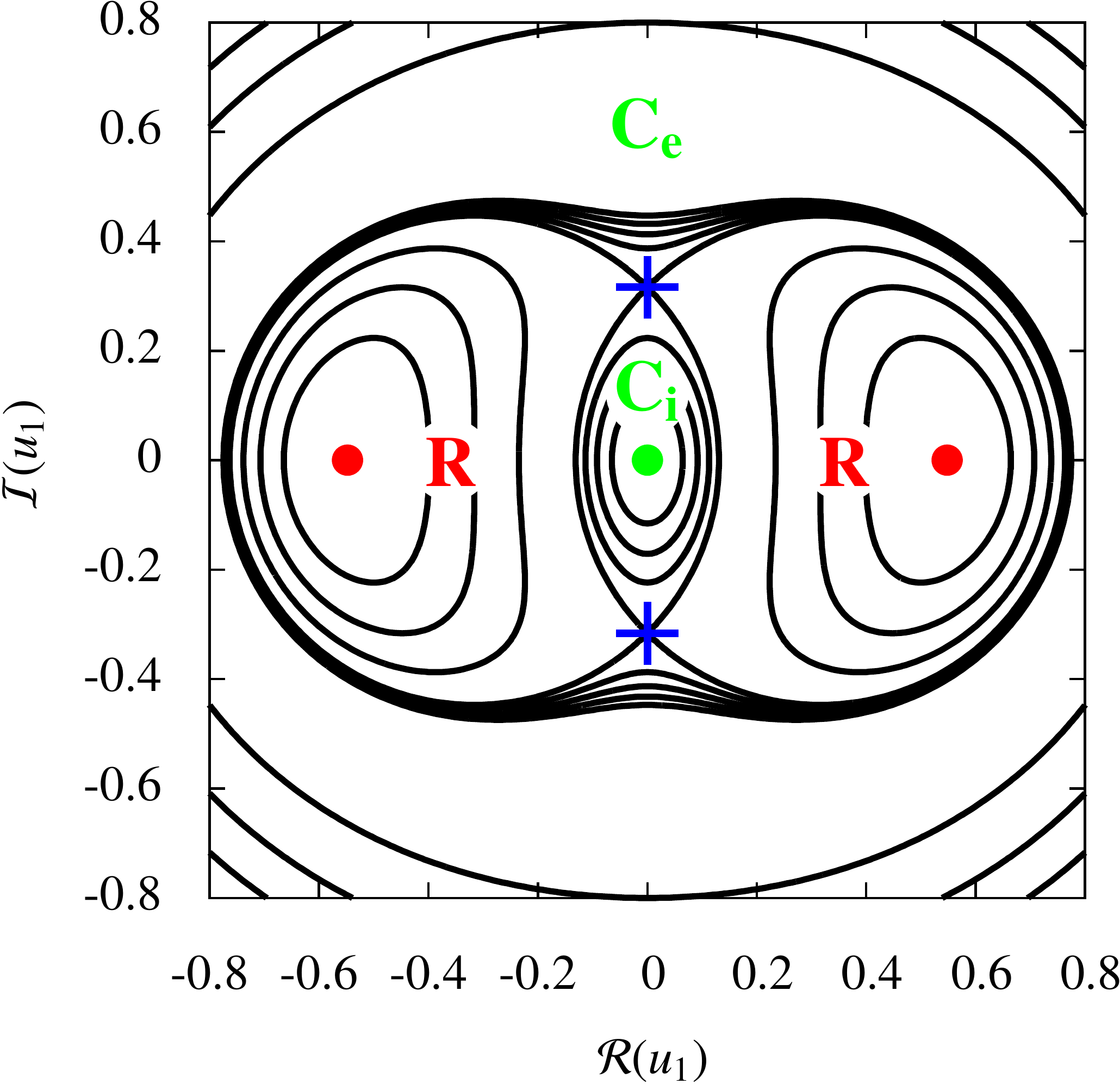}
    \caption{The three typical phase spaces of the simplified model (Hamiltonian~(\ref{eq:Hqd}))
      for a second order resonance. We plot the energy levels in the complex plane (real and
      imaginary parts of $u_1$), for $R = 0.1$ and $\delta$ = -0.2 (top), 0 (middle), 0.2 (bottom).
      Elliptical (stable) fixed points are marked with dots while hyperbolic (unstable) ones
      are marked with crosses.
      $\mathbf{R}$ corresponds to a resonant area and $\mathbf{C_e}$
      (respectively $\mathbf{C_i}$) to external (respectively internal) circulation.}
    \label{fig:II}
  \end{figure}
}

\newcommand\figIII{
  \begin{figure}
    \centering
    \includegraphics[width=9cm]{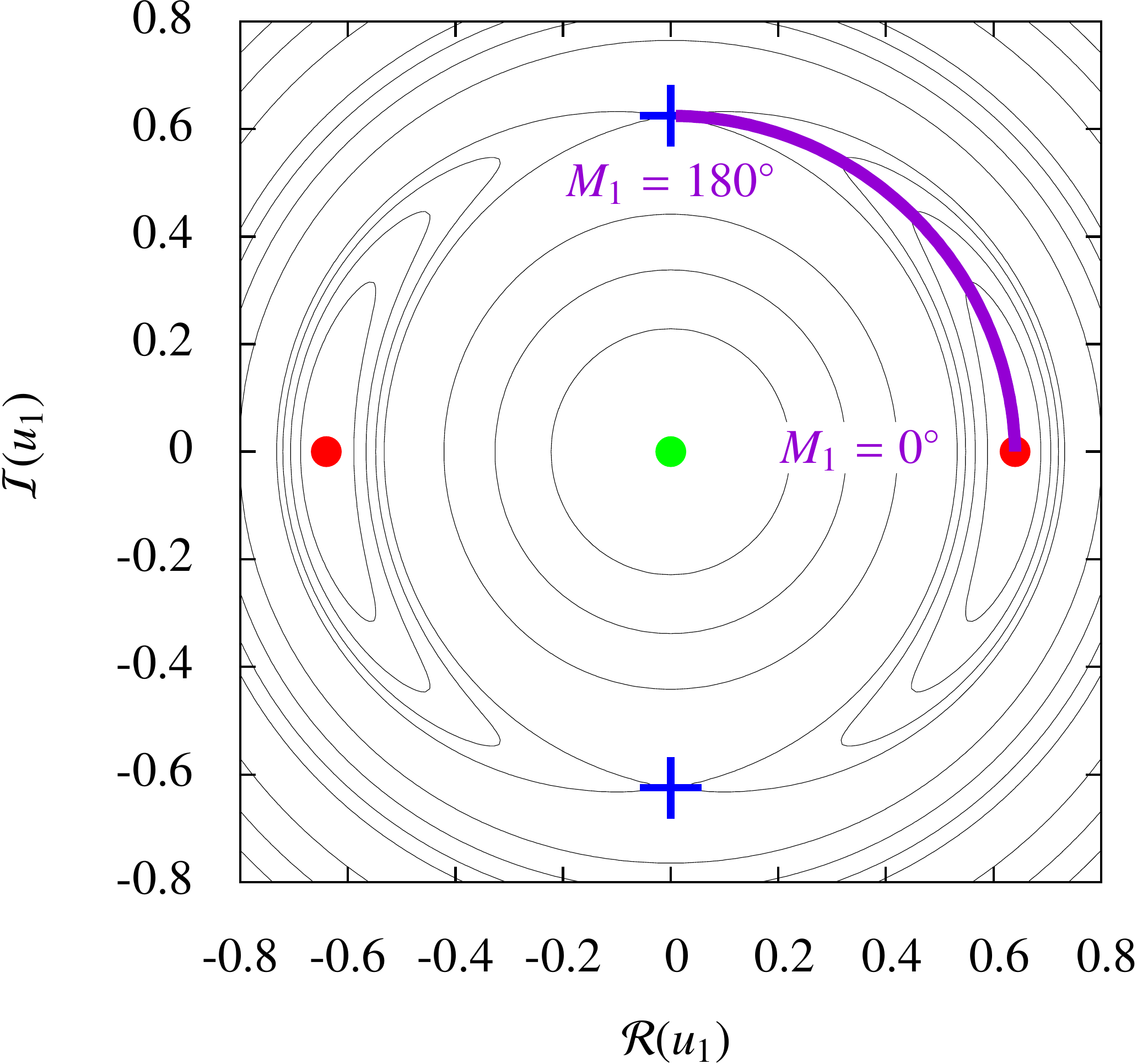}
    \caption{Diagram of initial conditions used in numerical simulations of \object{GJ~163}
      (3:1 MMR).
      $M_1$ varies between $0^\circ$ (center of libration) and $180^\circ$
      (separatrix of the resonance) along the purple line.}
    \label{fig:III}
  \end{figure}
}

\newcommand\figIV{
  \begin{figure*}
    \centering
    \includegraphics[width=8.5cm,trim = 0cm 2cm 0cm 0cm, clip]{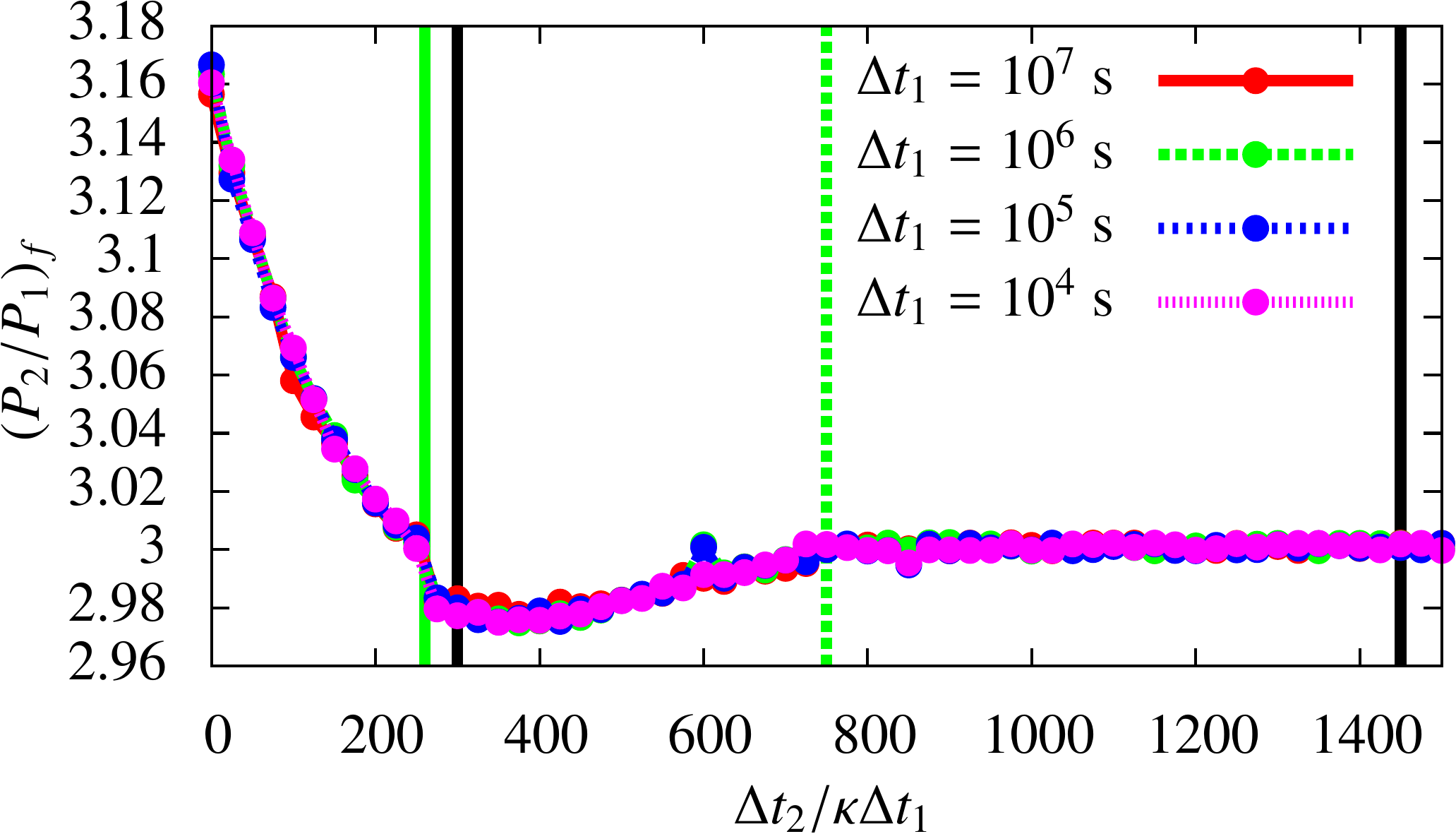} \hspace{5mm}
    \includegraphics[width=8.5cm,trim = 0cm 2cm 0cm 0cm, clip]{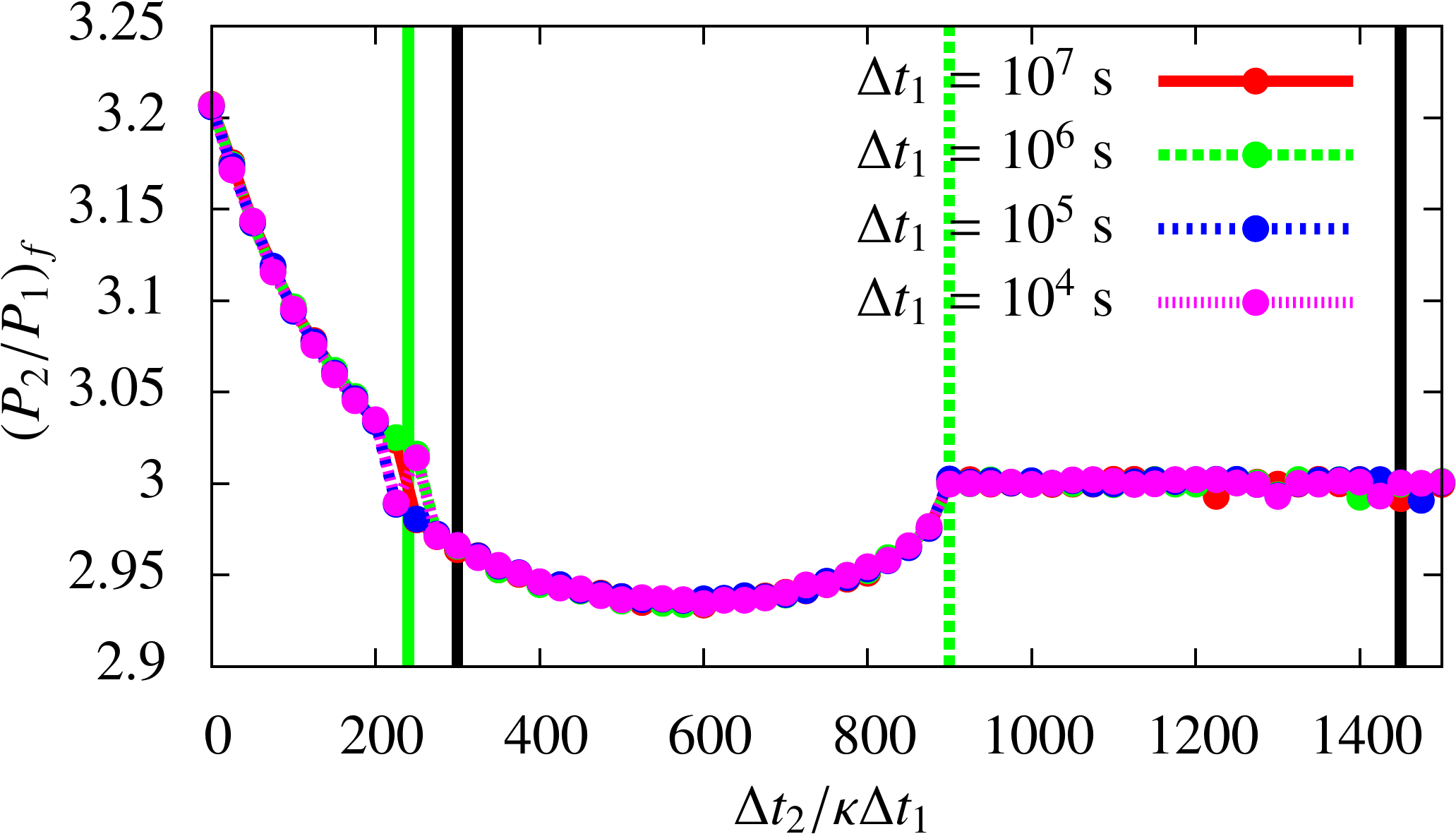}\\
    \includegraphics[width=8.5cm]{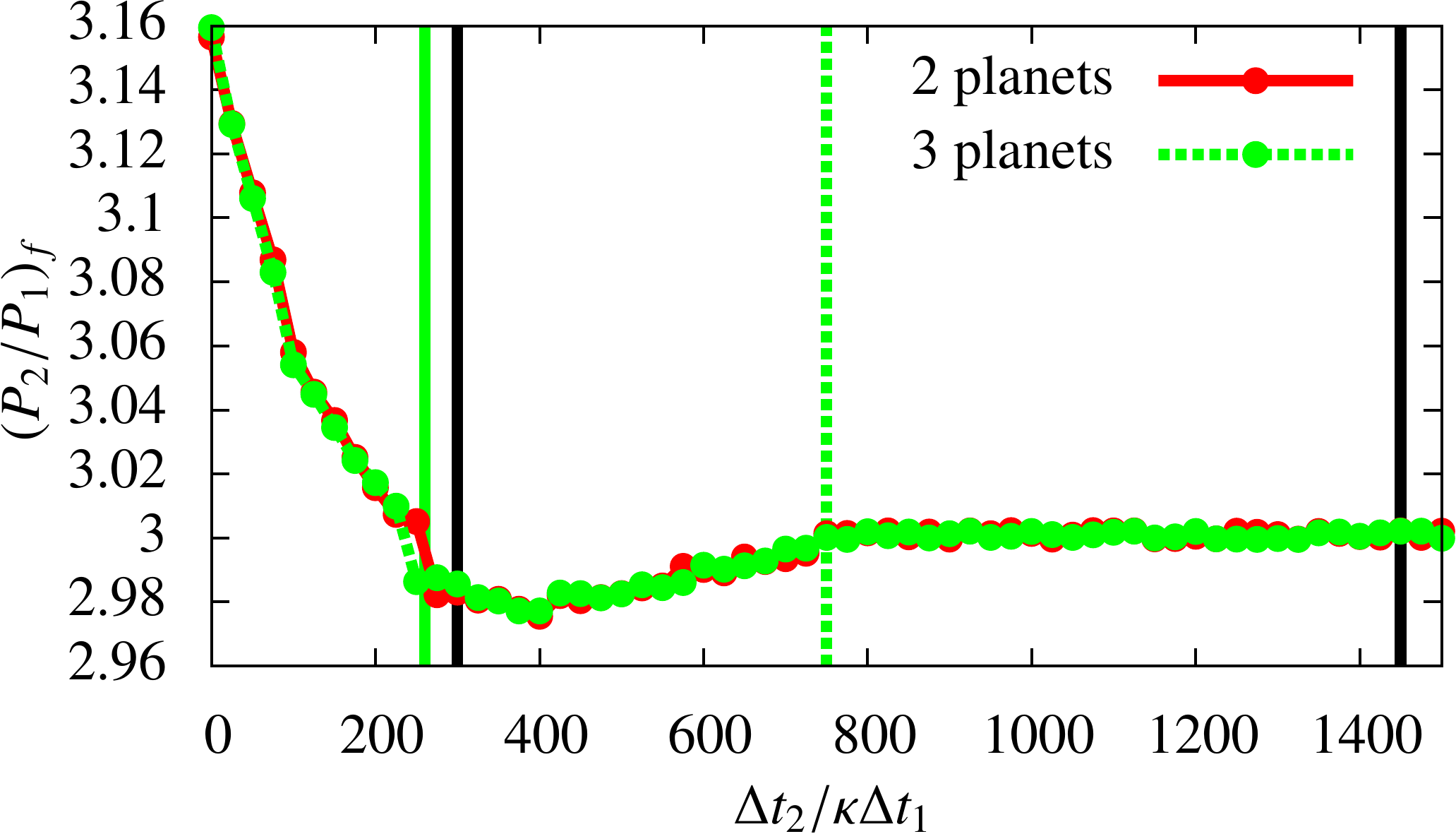} \hspace{5mm}
    \includegraphics[width=8.5cm]{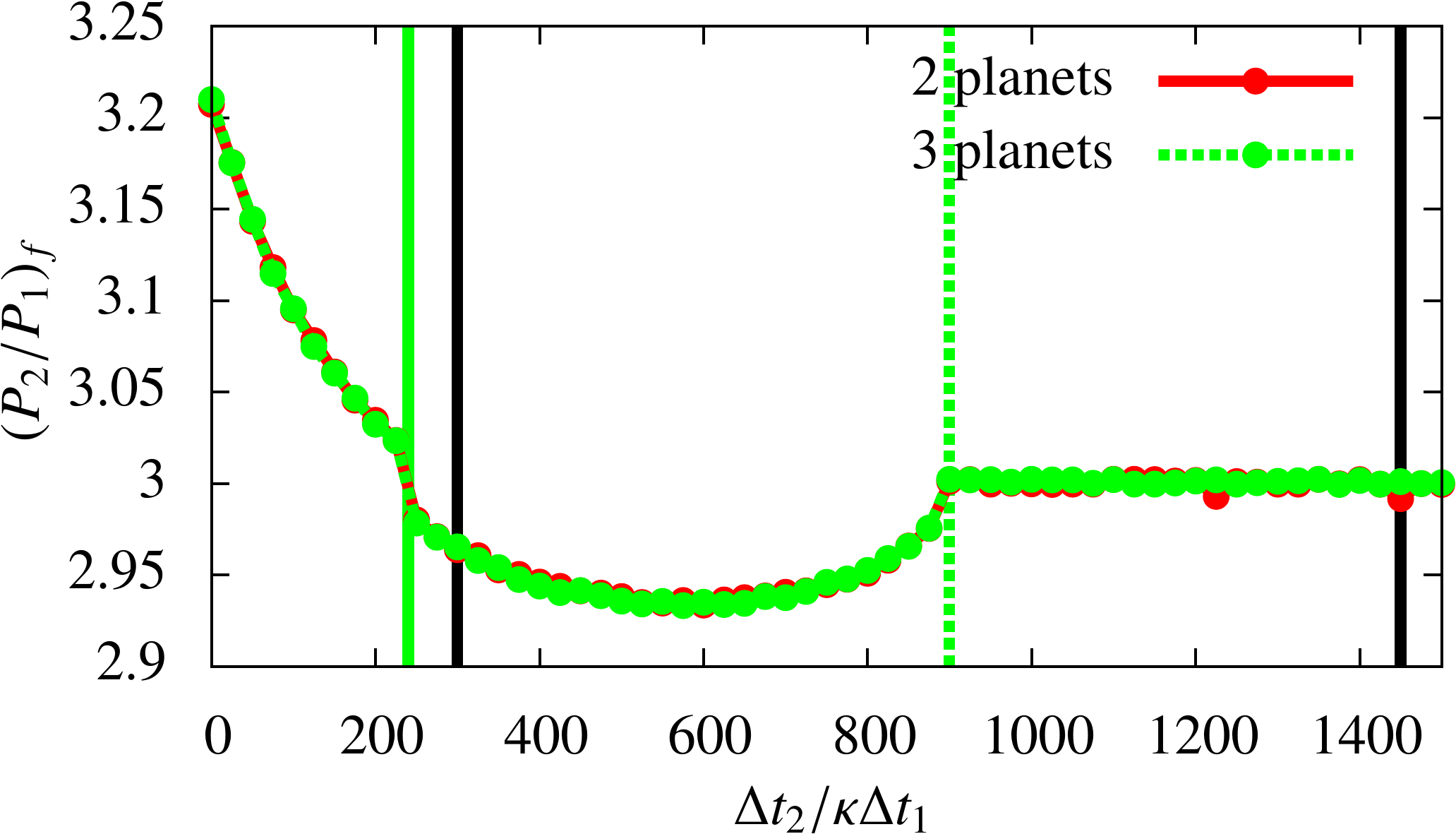}
    \caption{Final period ratio of \object{GJ~163}'s inner planets as a function of the dissipation
      balance between both planets \modif{($\Delta t_2/\kappa\Delta t_1$)}.
      We fix $M_1 = 100^\circ$ (left), $140^\circ$ (right)
      and compare the results obtained with $\Delta t_1 = 10^7$, $10^6$, $10^5$, and $10^4$ s (top)
      with an integration time of respectively 0.1, 1, 10, and 100 Myr which roughly corresponds
      to 10 Gyr for a realistic value of $\Delta t_1$ (100 s).
      We also compare these results with integrations taking into account the third planet that
      has been detected in the system \citep[bottom, see][]{bonfils_harps_2013}.
      For this latter comparison we use $\Delta t_1 = 10^7$ s.
      We observe only small variations when changing the timescale of the dissipation
      ($\Delta t_1$) and no systematic trend.
      The third planet does not seem to have an effect neither.
      The vertical black lines mark the range of \modif{$\Delta t_2/\kappa\Delta t_1$}
      that may conduct to internal circulation according to our analytical model.
      The green lines highlight the same range obtained with the simulations.
      The lower bound (green solid line) is at  $\Delta t_2/\kappa\Delta t_1 = 260$
      in the case $M_1 = 100^\circ$ (left) and 240 for $M_1 = 140^\circ$ (right).
      It does not vary much between both experiments and it is close to the analytical value (300).
      The upper bound (green dashed line) is at 750 (left) and 900 (right) while the analytical
      value is 1450.
      The difference between both numerical results, and between numerical and analytical values
      can easily be explained (see sect.~\ref{sec:n-body-simulations}).}
    \label{fig:IV}
  \end{figure*}
}

\newcommand\figV{
  \begin{figure*}
    \centering
    \includegraphics[width=6cm]{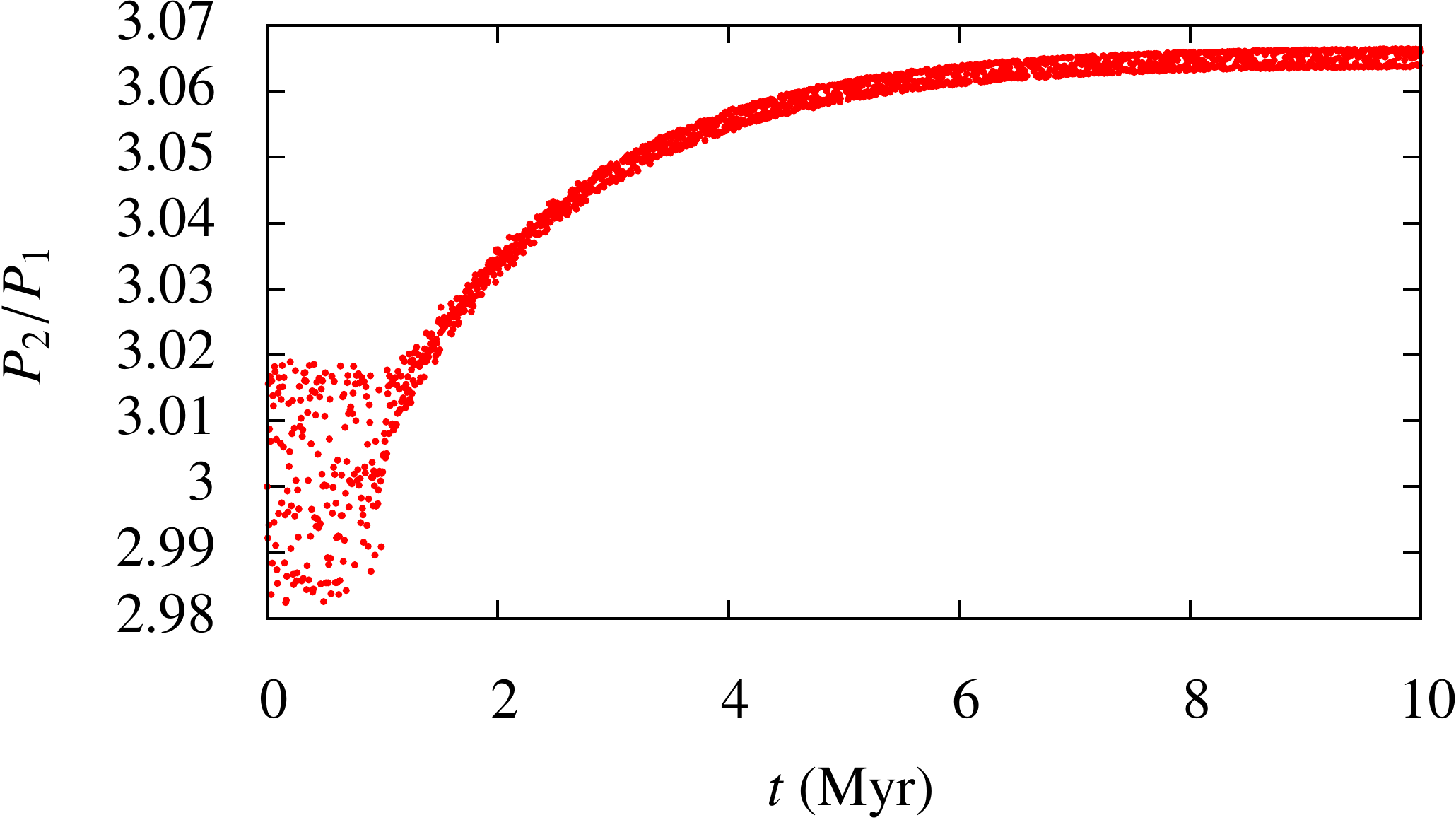}\hspace{1mm}
    \includegraphics[width=6cm]{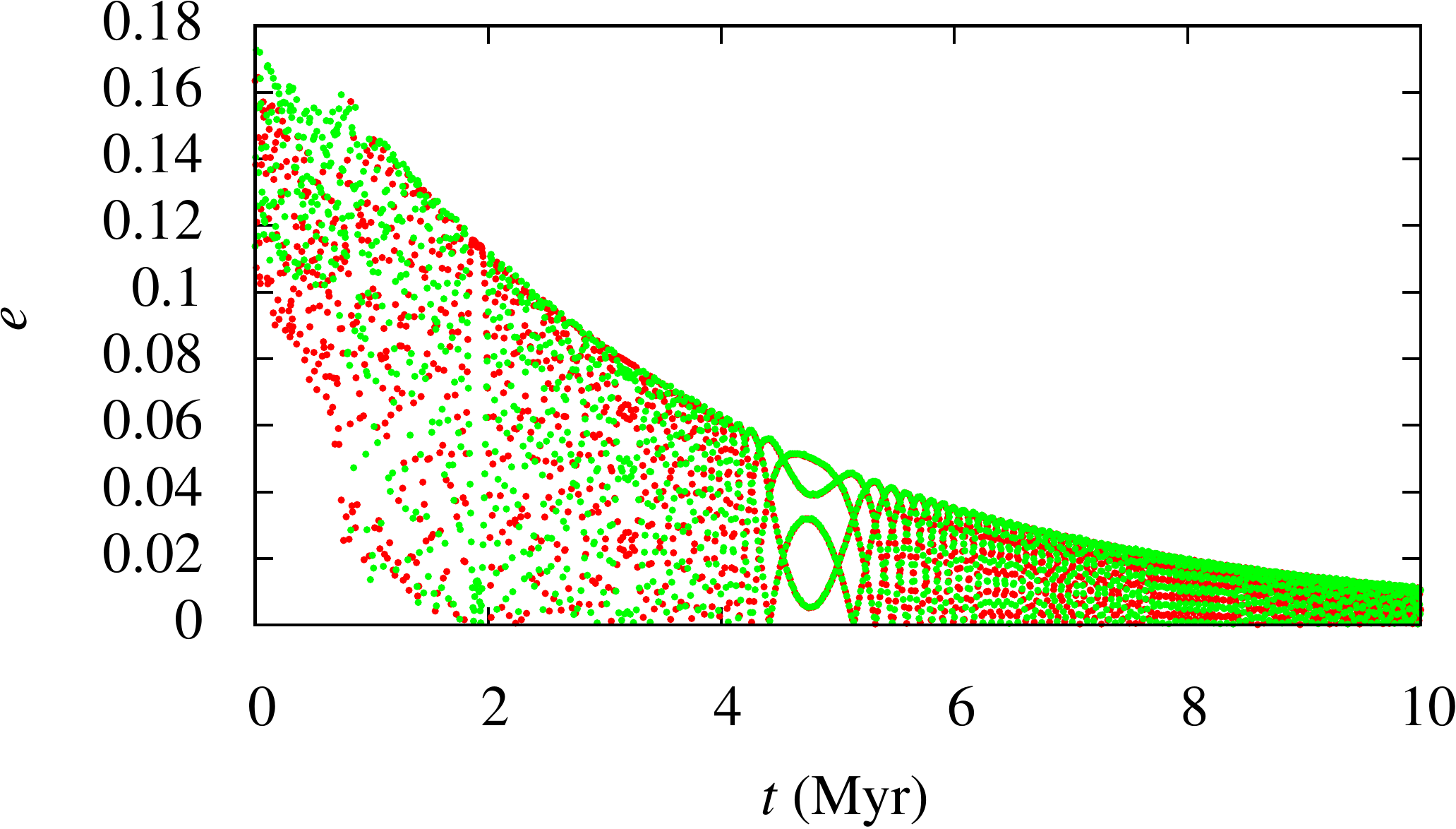}\hspace{1mm}
    \includegraphics[width=6cm]{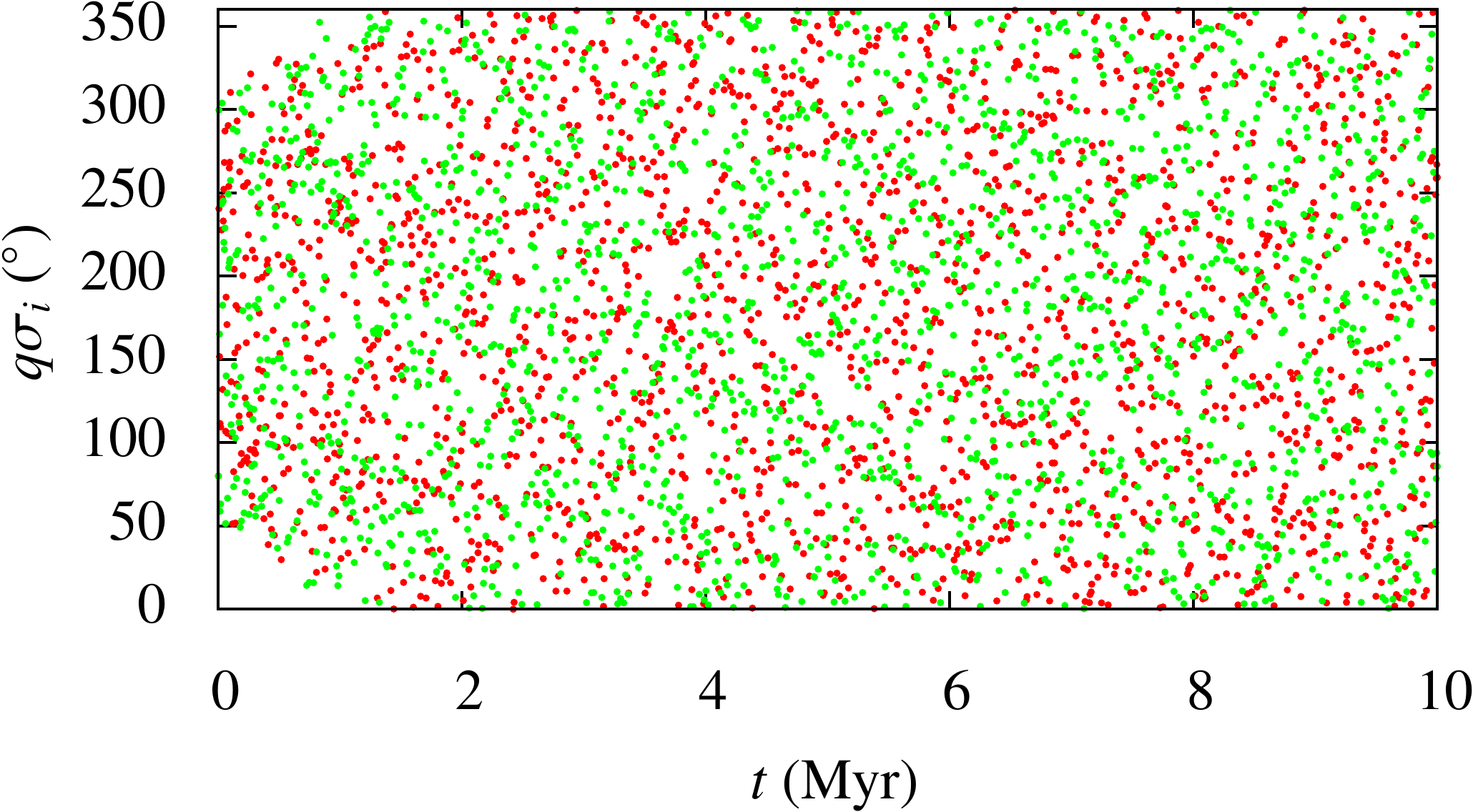}\\
    \includegraphics[width=6cm]{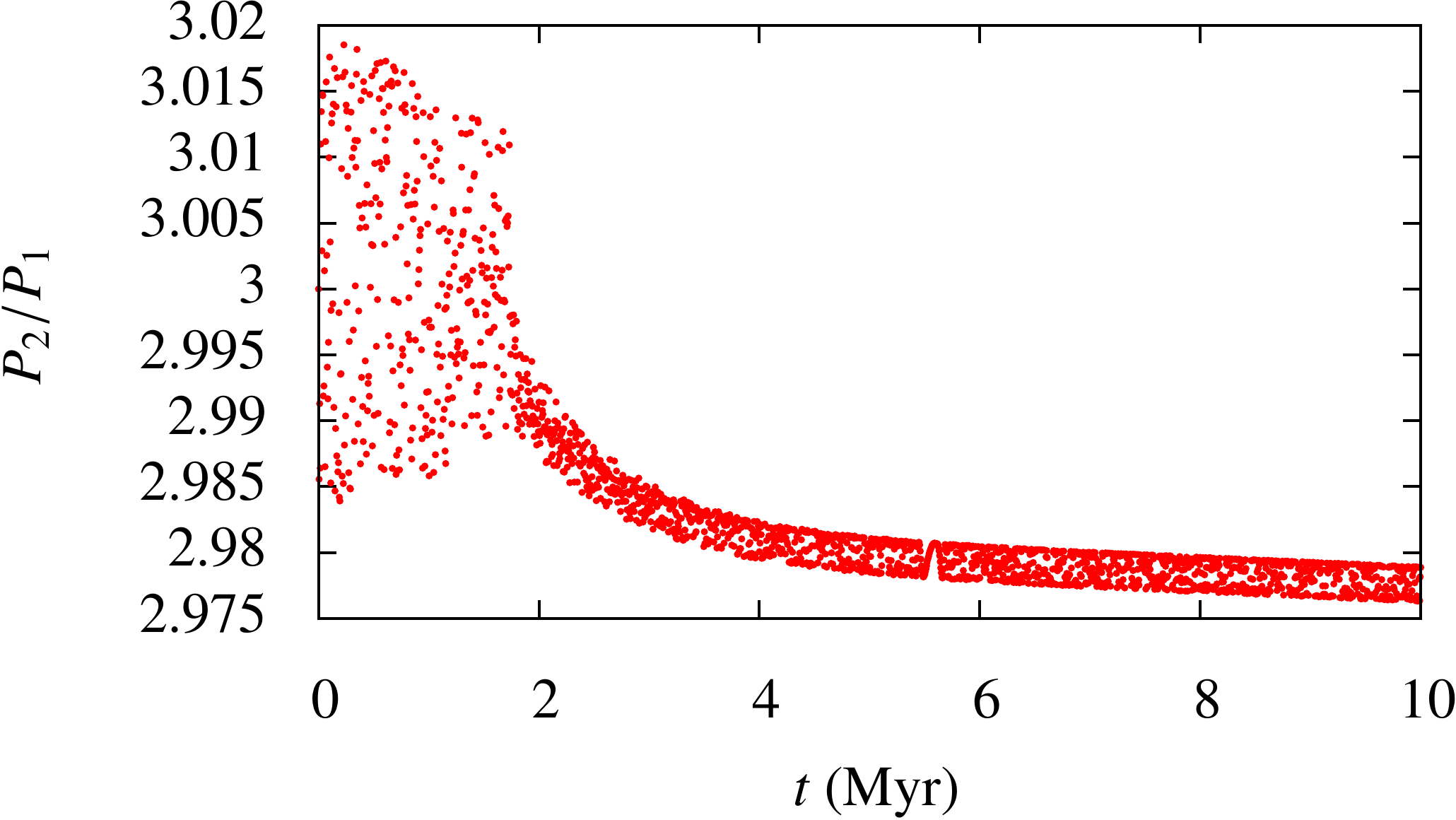}\hspace{1mm}
    \includegraphics[width=6cm]{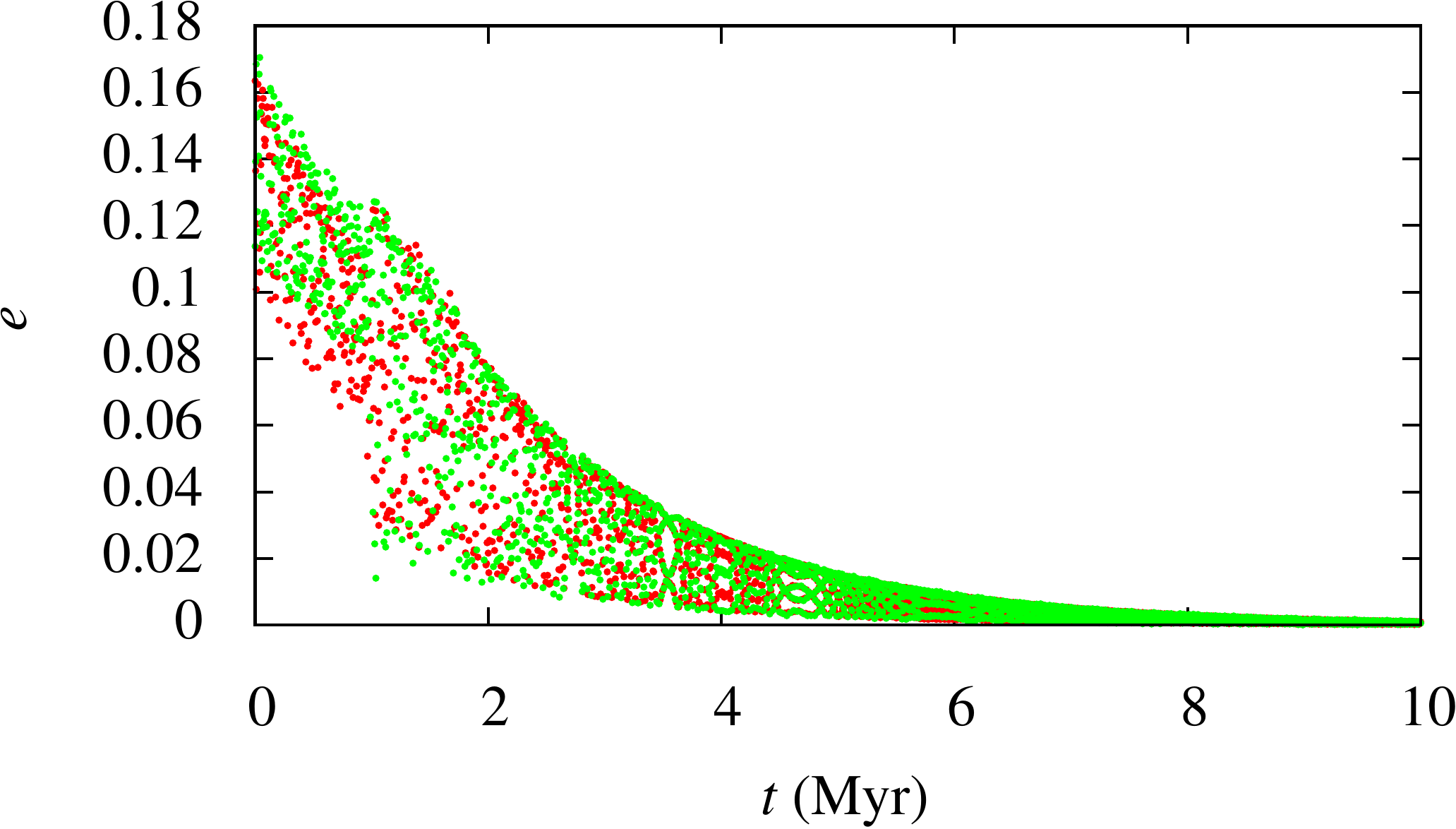}\hspace{1mm}
    \includegraphics[width=6cm]{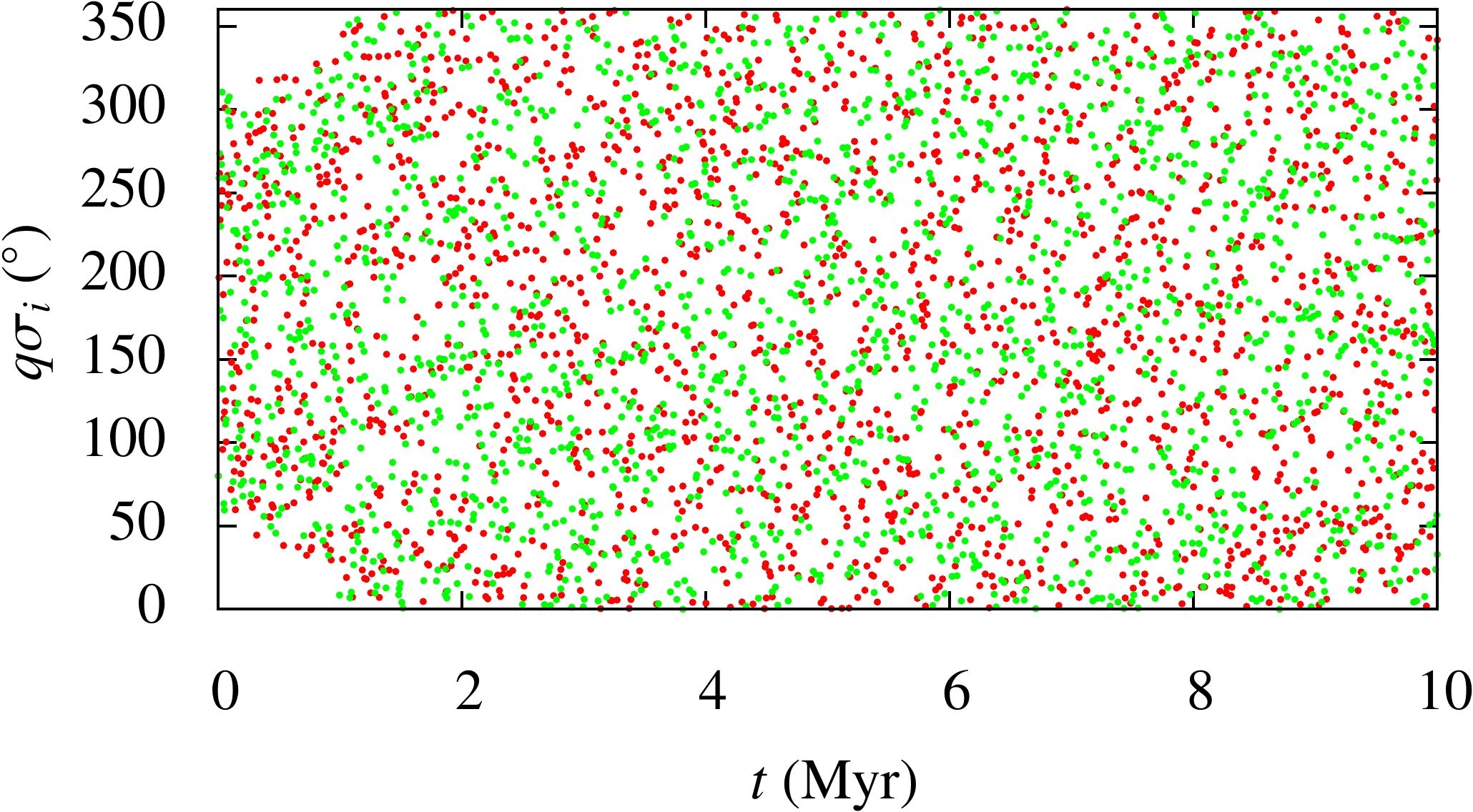}\\
    \includegraphics[width=6cm]{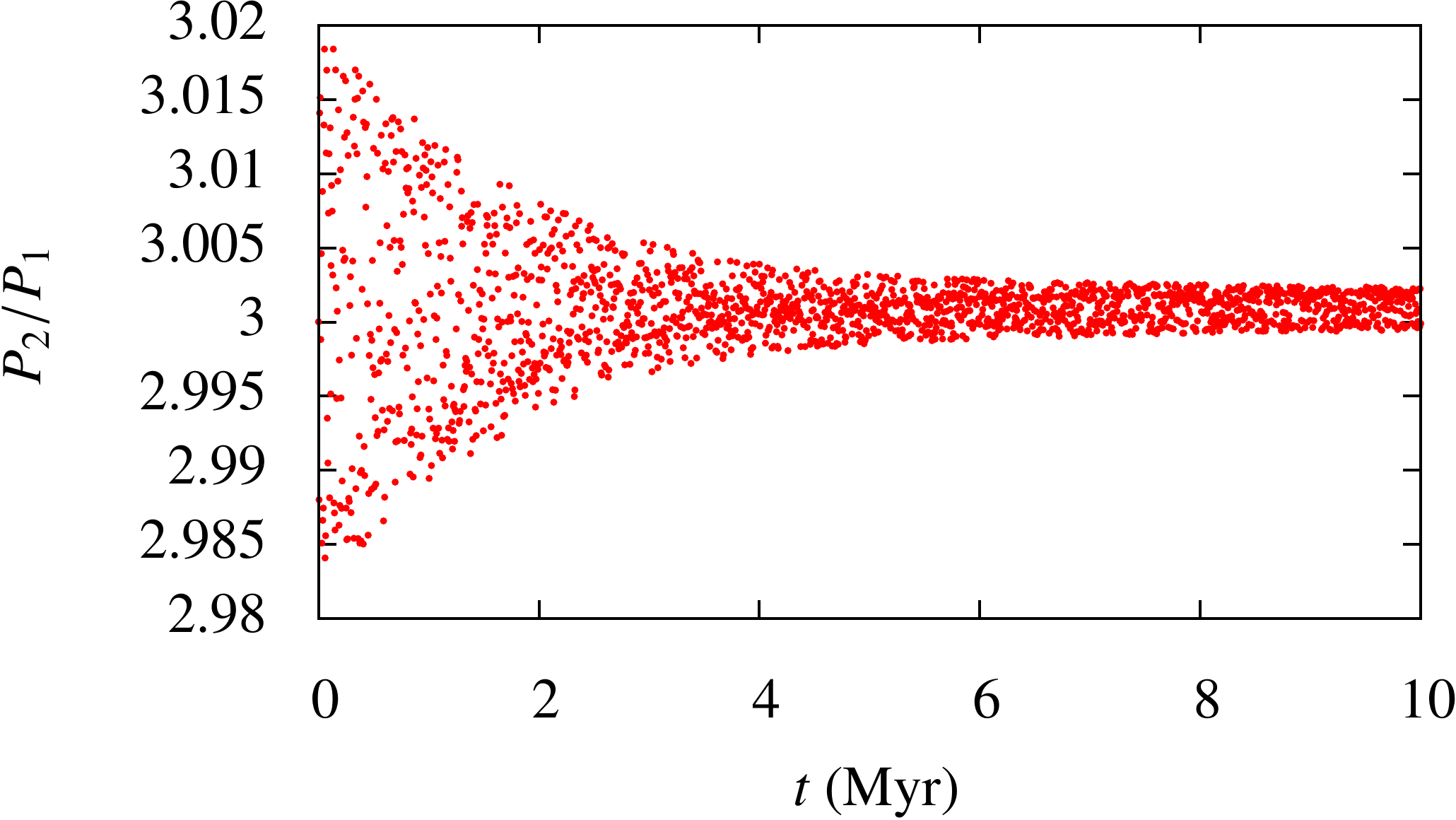}\hspace{1mm}
    \includegraphics[width=6cm]{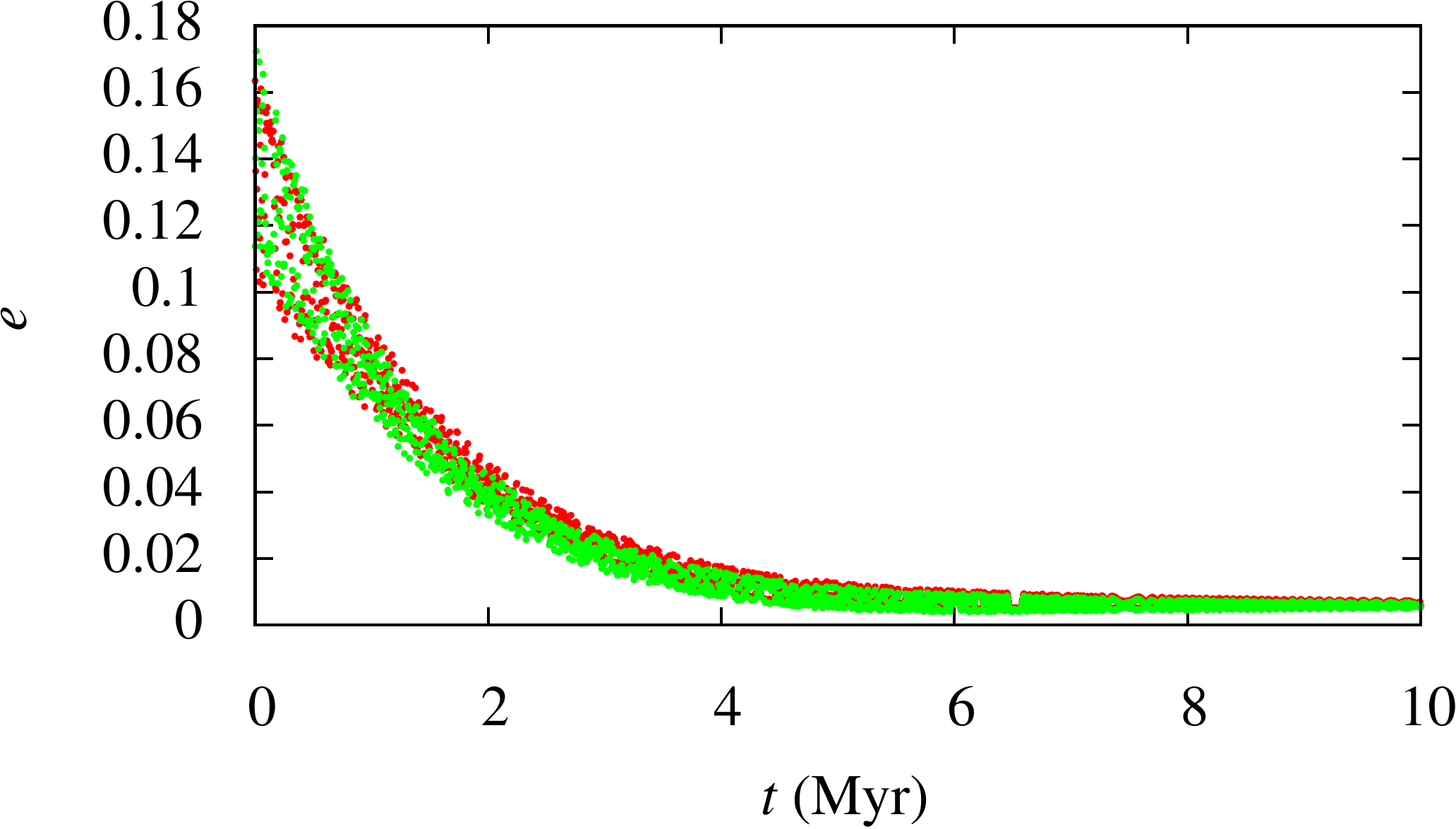}\hspace{1mm}
    \includegraphics[width=6cm]{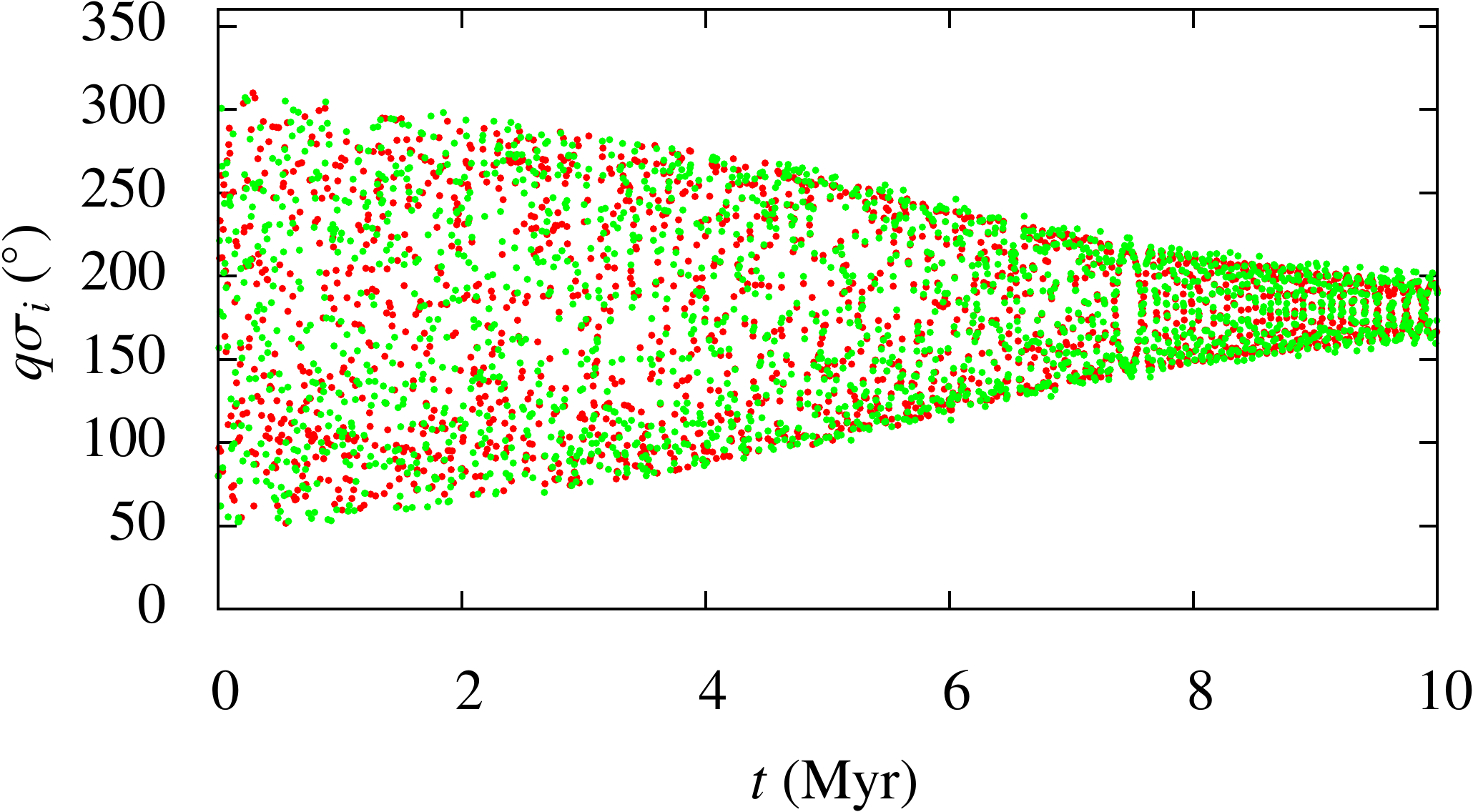}
    \caption{Evolution of the period ratio (left), the eccentricities (middle),
      and the resonant angles (right) of \object{GJ~163}'s inner planets
      for $M_1$ initially set to $100^\circ$ and \modif{$\Delta t_2/\kappa\Delta t_1 =$}
      100 (top), 400 (middle), and 1000 (bottom).
      Time is given in Myr but $\Delta t_1 = 10^5$ s in all these simulations.
      For a more realistic dissipation ($\Delta t_1 = 100$ s), the time should be read as Gyr.}
    \label{fig:V}
  \end{figure*}
}

\newcommand\figVI{
  \begin{figure}
    \centering
    \includegraphics[width=9cm]{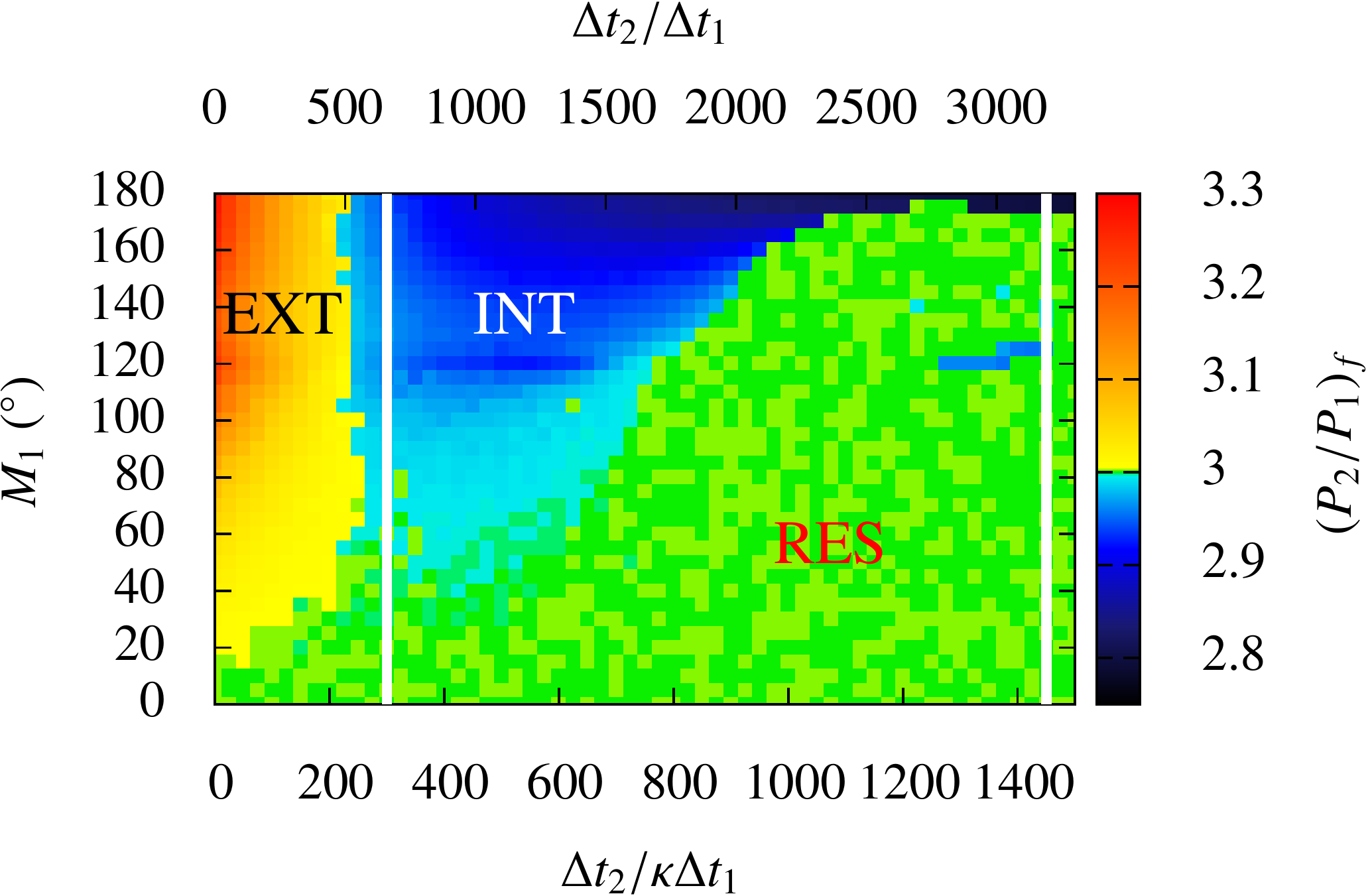}
    \caption{Final period ratio of \object{GJ~163}'s inner planets (3:1 MMR)
      as a function of the dissipation
      balance between both planets \modif{($\Delta t_2/\kappa\Delta t_1$, bottom x-axis)}
      and initial amplitude of libration ($M_1$).
      \modif{The coefficient $\kappa = k_{2,1}/k_{2,2}(R_1/R_2)^5$
        is unknown but can be estimated using a mass-radius power law.
        The top x-axis ($\Delta t_2/\Delta t_1$) is scaled using the value $\kappa \approx 2.2$
        obtained in Eq.~(\ref{eq:rhogj163})}
      When $M_1$ is set to $0^\circ$, the system begins at the libration
      center whereas when $M_1$ is set to $180^\circ$ it starts at the separatrix.
      We fixed $\Delta t_1 = 10^7$ s in order to speed-up the simulations
      and integrated the system during 0.1 Myr which roughly corresponds
      to 10 Gyr for a realistic value of $\Delta t_1$ (100 s).
      The vertical white lines mark the range of \modif{$\Delta t_2/\kappa\Delta t_1$}
      that may conduct to internal circulation according to our analytical model.
    }
    \label{fig:VI}
  \end{figure}
}

\newcommand\figVII{
  \begin{figure}
    \centering
    \includegraphics[width=9cm]{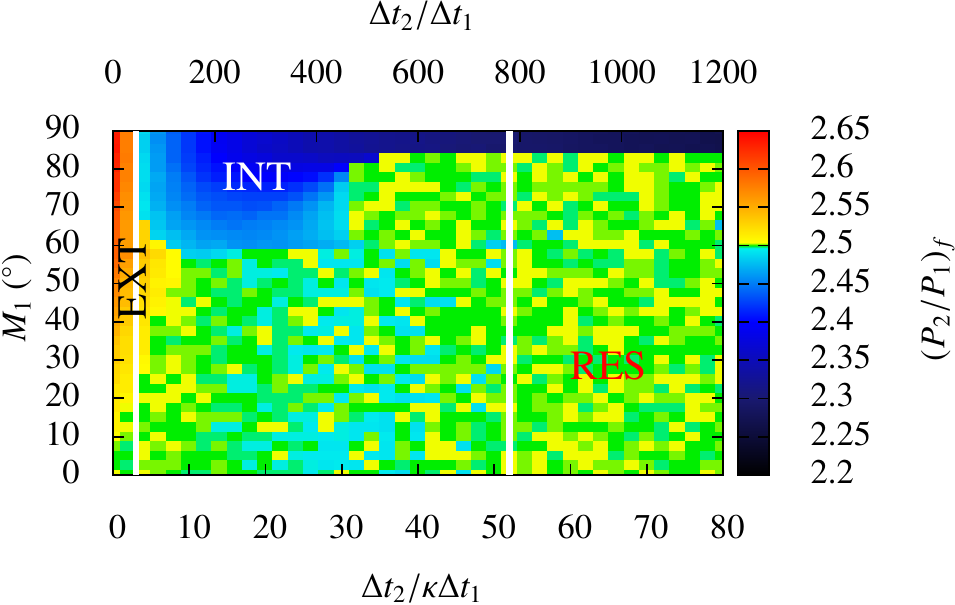}
    \caption{Same as Fig.~\ref{fig:VI} but for \object{GJ~581}b, and c (5:2 MMR).
      \modif{The top x-axis ($\Delta t_2/\Delta t_1$) is scaled using the value $\kappa \approx 15$
        obtained in Eq.~(\ref{eq:rhogj581}).}
      Note that $M_1$ varies only between $0^\circ$ (center of libration) and $90^\circ$
      (separatrix), because the resonant combination is $5 M_2 - 2 M_1$
      (factor 2 in front of $M_1$).}
    \label{fig:VII}
  \end{figure}
}

\newcommand\figVIII{
  \begin{figure}
    \centering
    \includegraphics[width=9cm]{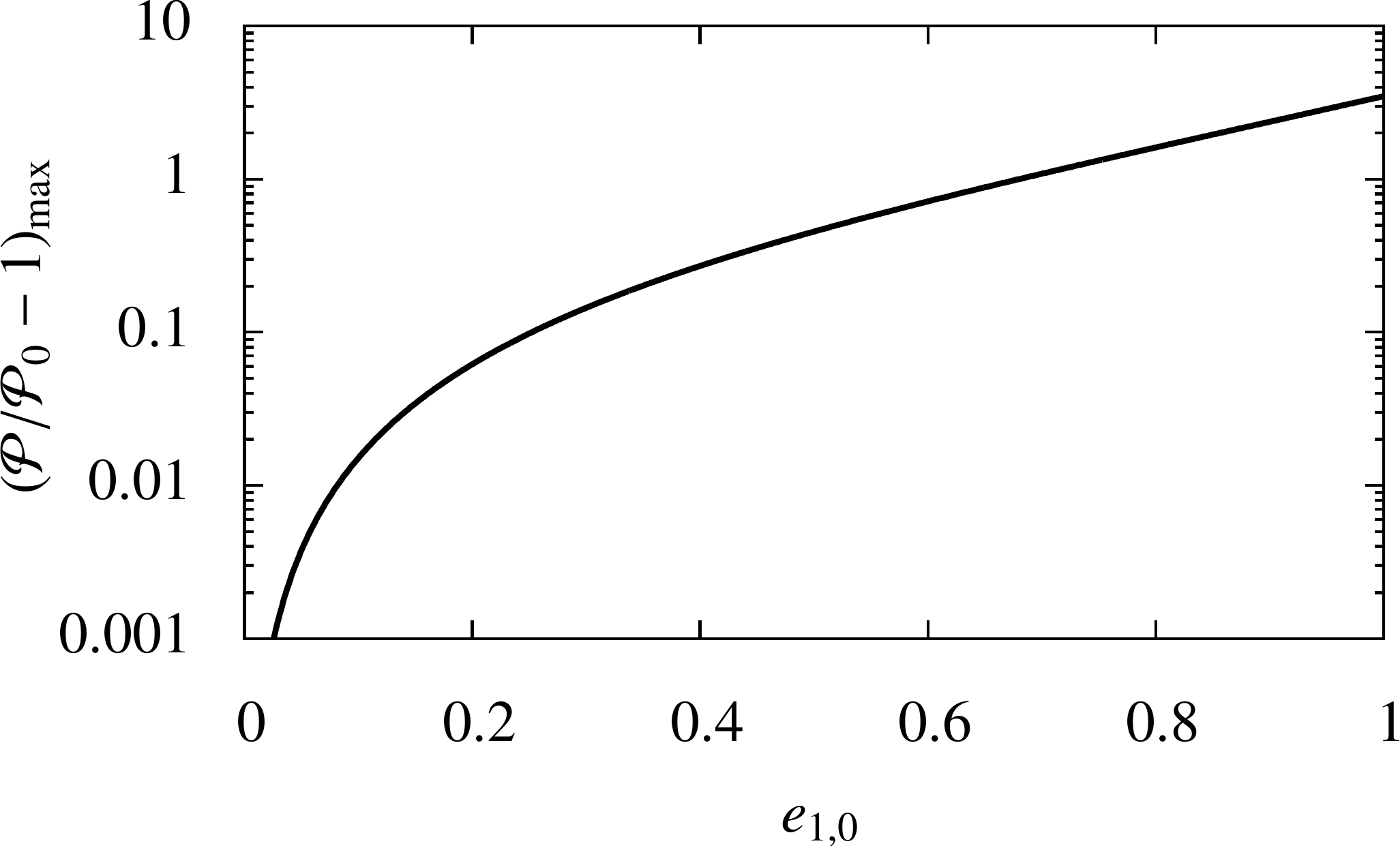}
    \caption{Maximum value reached by the period ratio $\mathcal{P} = P_2/P_1$ (when $t\to\infty$)
      as a function of the inner planet eccentricity
      when the system leaves the resonance ($e_{1,0}$).}
    \label{fig:VIII}
  \end{figure}
}

\newcommand\figIX{
  \begin{figure}
    \centering
    \includegraphics[width=9cm]{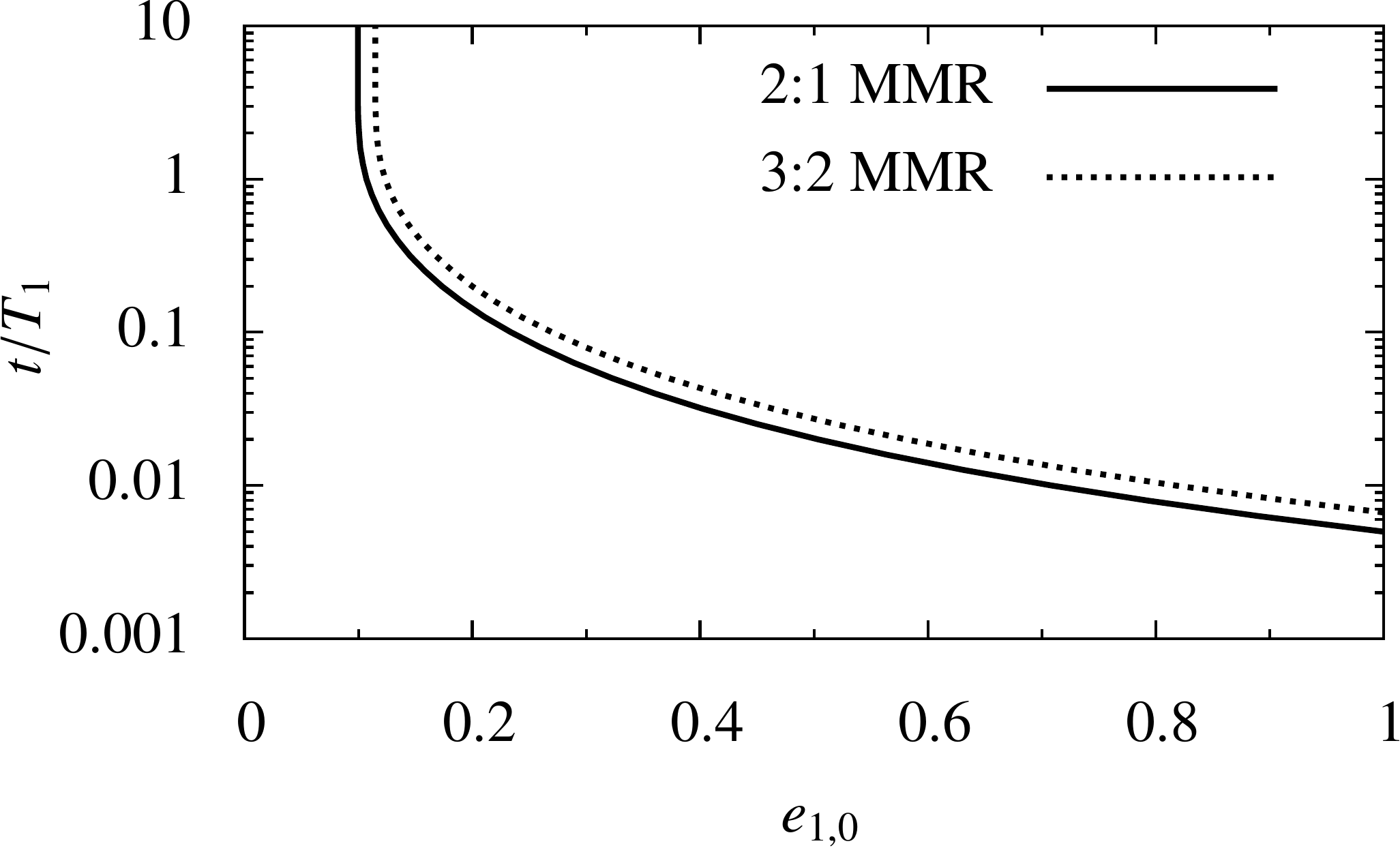}
    \caption{Time required to produce a departure of $P_2/P_1$ of 0.03 from the resonant value
      as a function of the eccentricity of the inner planet when the system leaves the resonance.
      This has to be compared with the estimate by \citet{lee_kepler_2013} considering a scenario
      of departure from the resonance at very low eccentricities ($t/T_1 \gtrsim 50-1000$).}
    \label{fig:IX}
  \end{figure}
}

\newcommand\tabI{
  \begin{table*}
    \begin{center}
      \caption{Orbital parameters of the \object{GJ~163} system \citep[from][fit with tidal constraint]{bonfils_harps_2013},
        and the \object{GJ~581} system \citep[from][]{forveille_only_2011}.
        The stellar masses are $0.4 \pm 0.04 M_\sun$ (\object{GJ~163}),
        and $0.31 \pm 0.02 M_\sun$ (\object{GJ~581}).
        Planets considered in this study are marked with bold font.}
      \small
      \begin{tabular}{cc|ccc|cccc}
        \hline
        & & \multicolumn{3}{c|}{\object{GJ~163}} & \multicolumn{4}{c}{\object{GJ~581}}\\
        Parameter & [unit] & \textbf{b} & \textbf{c} & d & e & \textbf{b} & \textbf{c} & d\\
        \hline
        $m \sin i$ & [$M_\earth$] &
        $\mathbf{10.661}$ & $\mathbf{7.263}$ & $22.072$ &
        $1.95$ & $\mathbf{15.86}$ & $\mathbf{5.34}$ & $6.06$\\
        $P$ & [days] &
        $\mathbf{8.633}$ & $\mathbf{25.645}$ & $600.895$ &
        $3.14945 \pm 0.00017$ & $\mathbf{5.36865 \pm 0.00009}$ & $\mathbf{12.9182 \pm 0.0022}$ & $66.64 \pm 0.08$\\
        $a$ & [AU] &
        $\mathbf{0.06069}$ & $\mathbf{0.12540}$ & $1.02689$ &
        $0.028$ & $\mathbf{0.041}$ & $\mathbf{0.073}$ & $0.22$\\
	$e$ & &
        $\mathbf{0.0106}$ & $\mathbf{0.0094}$ & $0.3990$ &
        $0.32 \pm 0.09$ & $\mathbf{0.031 \pm 0.014}$ & $\mathbf{0.07 \pm 0.06}$ & $0.25 \pm 0.09$ \\
        \hline
      \end{tabular}
      \label{tab:I}
    \end{center}
  \end{table*}
}

\begin{document}

\title{Resonance breaking due to dissipation in planar planetary systems}
\titlerunning{Resonance breaking due to dissipation}
\author{J.-B. Delisle\inst{1}
  \and J. Laskar\inst{1}
  \and A. C. M. Correia\inst{1,2}}

\institute{ASD, IMCCE-CNRS UMR8028, Observatoire de Paris, UPMC,
  77 Av. Denfert-Rochereau, 75014~Paris, France\\
  \email{delisle@imcce.fr}
  \and Departamento de F\'isica, I3N, Universidade de Aveiro, Campus Universit\'ario de Santiago,
  3810-193 Aveiro, Portugal
}

\date{\today}

\abstract{
  We study the evolution of two planets around a star,
  in mean-motion resonance and undergoing tidal effect.
  We derive an integrable analytical model of mean-motion resonances
  of any order which reproduce the main features of the resonant dynamics.
  \modif{
    Using this simplified model, we obtain a criterion showing that depending on the balance of the
    tidal dissipation in both planets,
    their final period ratio may stay at the resonant value, increase above, or decrease
    below the resonant value.}

  \modif{Applying this criterion to} the two inner planets orbiting \object{GJ~163},
  we deduce that the current period ratio (2.97) could be the outcome of
  dissipation in the 3:1 MMR provided that the innermost planet is gaseous
  (slow dissipation) while the second one is rocky (faster dissipation).
  We perform N-body simulations with tidal dissipation to confirm
  the results of our analytical model.

  \modif{We also apply our criterion on \object{GJ~581}b, c (5:2 MMR) and
    reproduce the current period ratio (2.4)
    if the inner planet is gaseous and the outer is rocky (as for \object{GJ~163}).}

  Finally, we apply our model to the \textit{Kepler} mission's statistics.
  We show that the excess of planets pairs close to first order MMR but in external circulation,
  i.e., with period ratios $P_{out}/P_{in} > (p+1)/p$ for the resonance $(p+1)$:$p$,
  can be reproduced by tidal dissipation in the inner planet.
  \modif{There is no need for any other dissipative mechanism,
  provided that these systems left the resonance with non-negligible eccentricities.}
}


\maketitle

\section{Introduction}
\label{sec:introduction}
It has been shown that planets in first order mean-motion resonances (MMR) that
undergo tidal dissipation naturally leave the resonant configuration by moving away from each other
\citep{papaloizou_dynamics_2010,papaloizou_tidal_2011,lithwick_resonant_2012,delisle_dissipation_2012,batygin_dissipative_2013}.
The tidal dissipation first induces a decrease of both eccentricities (as expected) and the system
initially stays in resonance.
However, when eccentricities reach low values, the ratio between the orbital periods
of the outer planet and the inner one begins to increase (diverging orbits)
as the eccentricities continue to decrease.
If the timescale of the dissipation is sufficiently short (compared to the age of the system),
the period ratio can significantly depart from the resonant value.

It is important to note that during this process, the system never crosses
the resonance separatrix.
Indeed, the separatrix simply disappears at low eccentricities,
and the system end-up with a period ratio $P_{out}/P_{in}$ greater than the resonant value
\citep[e.g.][]{delisle_dissipation_2012}.
However, if the amplitude of libration in the resonance becomes sufficiently high, the system
may cross the separatrix before it disappears and may
end-up either in the internal or the external circulation areas
\citep[e.g.][]{novak_interesting_2003,goldreich_overstable_2014}.
External circulation refers to the configuration where planets are close to a MMR $(p+q)$:$p$,
but with a period ratio greater than the resonant value ($P_{out}/P_{in}>(p+q)/p$).
On the contrary, internal circulation refers to the configuration $P_{out}/P_{in}<(p+q)/p$.

In this study, we obtain a simple criterion on the dissipation undergone by planetary
systems in MMR of any order to end-up in the resonant area, the internal or the external circulations areas.
In section~\ref{sec:an-integr-simpl} we present a method to obtain a simple integrable model
for resonances of any order.
In section~\ref{sec:diss-reson-syst} we consider the evolution of the dynamics under dissipation.
We focus on the evolution of the amplitude of libration in the resonance depending on the balance
of dissipation in both planets.
In section~\ref{sec:applications} we apply our reasoning to
\object{GJ~163} (3:1 MMR) and \object{GJ~581} (5:2 MMR) planetary systems.
Finally, in section~\ref{sec:kepler} we discuss the impact of resonance breaking induced by tides
on \textit{Kepler} mission's multi-planetary systems statistics.

\section{An integrable simplified model of resonances}
\label{sec:an-integr-simpl}

In order to obtain simple criteria on the dissipation we need an analytic\modif{al} model of MMR as simple
as possible that still captures the main characteristics of the resonant dynamics.
In particular, the model should be integrable (one degree of freedom).
For first order resonances, such an integrable approximation is easily obtained
\citep[see][]{sessin_motion_1984,henrard_reducing_1986,
  wisdom_canonical_1986,batygin_analytical_2013}.
However, for higher order resonances, the situation is more complex and we need to make some
simplifying hypotheses.

First we introduce our notations and the classical method for obtaining a non-integrable Hamiltonian
model for any MMR.
Then, we reproduce the well known method to obtain an integrable one for first order resonances
as a template for higher order ones.

\subsection{Hamiltonian of a \modif{$(p+q)$:$p$} MMR}
Let us refer to the star as body 0, to the inner planet as body 1, and to the outer one as body 2.
Noting the masses of the three bodies $m_i$, we introduce for both planets
$\mu_i = \mathcal{G} (m_0 + m_i)$
and $\beta_i = m_0 m_i/(m_0 + m_i)$,
where $\mathcal{G}$ is the gravitational constant.
We note $\mathbf{r}_i$ the position vector of the planets with respect to the star and
$\tilde{\mathbf{r}}_i$ the canonically conjugated momenta
\citep[in astrocentric coordinates, see][]{laskar_stability_1995}.
As usual in the literature, semi-major axes are noted $a_i$, eccentricities $e_i$,
mean longitudes $\lambda_i$, longitudes of periastron $\varpi_i$.
It should be noted that we only consider the planar case in this study.

The Hamiltonian of the three body problem reads:
\begin{equation}
  \label{eq:H3b}
  \scaled{\mathcal{H}} = \scaled{\mathcal{K}} + \scaled{\mathcal{H}}_1
\end{equation}
where $\scaled{\mathcal{K}}$ is the Keplerian part (star-planets interactions)
and $\scaled{\mathcal{H}}_1$ is the perturbative part (planet-planet interactions).
The Keplerian part is given by:
\begin{equation}
  \label{eq:Kep}
  \scaled{\mathcal{K}} = - \sum_{i=1}^2 \frac{\mu_i^2 \beta_i^3}{2 \scaled{\Lambda}_i^2}
\end{equation}
where $\scaled{\Lambda}_i$ is the circular angular momentum of planet $i$:
\begin{equation}
  \scaled{\Lambda}_i = \beta_i \sqrt{\mu_i a_i}
\end{equation}

The perturbative part can be decomposed in direct and indirect interactions:

\begin{eqnarray}
  \label{eq:perturb}
  \scaled{\mathcal{H}}_1 &=& \scaled{\mathcal{U}}_1 + \scaled{\mathcal{T}}_1\\
  \scaled{\mathcal{U}}_1 &=& -\mathcal{G} \frac{m_1m_2}{\Delta_{12}}\\
  \scaled{\mathcal{T}}_1 &=& \modif{\frac{\mathbf{\tilde{r}}_1 . \mathbf{\tilde{r}}_2}{m_0}}
\end{eqnarray}

with $\Delta_{12} = ||\mathbf{r}_1-\mathbf{r}_2||$.

The perturbation can be expressed as a function of elliptical orbital elements by expanding it
in Fourier series of the angles $\lambda_i, \varpi_i$ \citep[e.g.][]{laskar_stability_1995}.

For a given mean-motion resonance $(p+q)$:$p$ between both planets, the corresponding
combination of mean-longitudes undergoes slow variations:
\begin{equation}
  (p+q)\dot\lambda_2 - p\dot\lambda_1 \approx 0
\end{equation}
whereas other (non-resonant) combinations of these angles are circulating rapidly.
The long term evolution of the orbits is well described by the averaged Hamiltonian
over the non-resonant combinations of the mean longitudes.

By performing this averaging and the classical angular momentum reduction,
one obtains a two degrees of freedom problem \citep[e.g.][]{delisle_dissipation_2012} and
two constant\modif{s} of motion.
The first constant of motion is the total angular momentum:
\begin{equation}
  \label{eq:G}
  \scaled{G} = \scaled{G}_1 + \scaled{G}_2 = \scaled{\Lambda}_1\sqrt{1-e_1^2}
  + \scaled{\Lambda}_2\sqrt{1-e_2^2}
\end{equation}

The second one, coming from the averaging, is a combination of the circular angular momenta
\modif{\citep[or semi-major axes, e.g.][]{michtchenko_modeling_2001}:}
\begin{equation}
  \label{eq:Gamma}
  \Gamma = \frac{p+q}{p} \scaled{\Lambda}_1 + \scaled{\Lambda}_2
\end{equation}
As shown in \citet{delisle_dissipation_2012}, the constant $\Gamma$ can be used as a scaling
factor and does not influence the dynamics of the system
except by changing the scales of the problem (in space, energy, and time).
The elimination of $\Gamma$ is achieved by performing the following change of coordinates
\citep[see][]{delisle_dissipation_2012}:
\begin{eqnarray}
  \Lambda_i &=& \frac{\scaled{\Lambda}_i}{\Gamma}\\
  G_i &=& \frac{\scaled{G}_i}{\Gamma}\\
  G &=& \frac{\scaled{G}}{\Gamma} = G_1 + G_2\\
  \mathcal{H} &=& \Gamma^2 \scaled{\mathcal{H}}\\
  t &=& \frac{\scaled{t}}{\Gamma^3}
\end{eqnarray}
while angle coordinates are unchanged.
Using these new coordinates, the dynamics of the system depends on only one parameter:
$G$, the renormalized angular momentum.

The remaining two degrees of freedom can be represented by both resonant angles:
\begin{equation}
  \label{eq:sigma}
  \sigma_i = -\frac{p}{q}\lambda_1 + \frac{p+q}{q} \lambda_2 - \varpi_i
\end{equation}
and both deficits of angular momentum \citep{laskar_spacing_2000} which are canonically conjugated to the resonant angles:
\begin{equation}
  \label{eq:AMDi}
  I_i = \Lambda_i - G_i
\end{equation}

We may also introduce rectangular coordinates:
\begin{equation}
  \label{eq:xi}
  x_i = \sqrt{I}_i \expo{\ImUnit \sigma_i}
\end{equation}
It should be noted that for small eccentricities, $|x_i| \propto e_i$.
The averaged Hamiltonian takes the form:
\begin{equation}
  \label{eq:Haver}
  \mathcal{H} = \mathcal{K}(I_i) + \mathcal{S}(I_i, \Delta\varpi)
  + \mathcal{R}(I_i,\sigma_i)
\end{equation}
where $\mathcal{S}$ is the secular part of the Hamiltonian depending on the difference
of longitudes of periastron ($\Delta\varpi=\sigma_2-\sigma_1$)
but not on mean longitudes of the planets
and $\mathcal{R}$ is the resonant part.
These two parts can be expanded as power series of eccentricities (more precisely of $x_i$).
The method used to obtain these expansion are presented in
\citet{laskar_stability_1995,delisle_dissipation_2012}.

The Keplerian part can be expressed as a function of the momenta $I_i$ by substituting
the expressions of $\Lambda_i$ in Equation~(\ref{eq:Kep}):
\begin{eqnarray}
  \label{eq:La1}
  \Lambda_1 &=& \frac{p}{q} \left[1 - (G + \mathcal{D}) \right]\\
  \label{eq:La2}
  \Lambda_2 &=& \frac{p+q}{q} \left(G + \mathcal{D} \right) - \frac{p}{q}
\end{eqnarray}
with $\mathcal{D}$, the total angular momentum deficit \citep{laskar_spacing_2000}:
\begin{equation}
  \mathcal{D} = I_1 + I_2 = \sum_{i=1}^2 x_i\bar{x}_i
\end{equation}

The secular part contains terms of degree two and more in eccentricities while the resonant part
contains terms of degree $q$ and more.
Thus, the simplest model of the resonance should take into account
at least those terms of order $q$ in eccentricities in the perturbative part:
\begin{equation}
  \label{eq:Hq}
  \mathcal{H} = \mathcal{K}(\mathcal{D}) + \mathcal{S}_q(I_i, \Delta\varpi)
  + \sum_{k=0}^q R_k (x_1^kx_2^{q-k} + \bar{x}_1^k\bar{x}_2^{q-k})
\end{equation}
where $\mathcal{S}_q$ is the secular part truncated at degree q and $R_k$ are
constant coefficient \citep[see][]{delisle_dissipation_2012}.

This problem is much simpler than the initial four degrees of freedom problem.
However, in general it is still non-integrable since it presents two degrees of freedom.

\subsection{First order resonances}
\label{sec:first-order-reson}

For a first order MMR (such as the 2:1 or 3:2 resonances) the simplest Hamiltonian reads:
\begin{equation}
  \label{eq:H1}
  \mathcal{H} = \mathcal{K}(\mathcal{D}) + R_1 (x_1 + \bar{x}_1) + R_0 (x_2 + \bar{x}_2)
\end{equation}
where there are no secular terms since they only appear at degree two.
It is well known that the Hamiltonian (\ref{eq:H1}) is integrable
\citep[see][]{sessin_motion_1984,henrard_reducing_1986,
  wisdom_canonical_1986,batygin_analytical_2013}.
Introducing $R$ and $\phi$ such as:
\begin{eqnarray}
  R_1 &=& R \cos(\phi)\\
  R_0 &=& R \sin(\phi)
\end{eqnarray}
and the new coordinates $u_1$, $u_2$ such as:
\begin{equation}
  x = R_\phi u
\end{equation}
with $R_\phi$ the rotation matrix:
\begin{equation}
  R_\phi =
  \left( \begin{array}{ c c }
      \cos\phi & -\sin\phi\\
      \sin\phi & \cos\phi\\
    \end{array} \right)
\end{equation}
the Hamiltonian (\ref{eq:H1}) reads:
\begin{equation}
  \label{eq:H1u}
  \mathcal{H} = \mathcal{K}(\mathcal{D}) + R (u_1 + \bar{u}_1)
\end{equation}
with:
\begin{equation}
  \mathcal{D} = \sum_{i=1}^2 u_i\bar{u}_i
\end{equation}

The Hamiltonian (\ref{eq:H1u}) does not depend on the angle associated with $u_2$
but only on the action $u_2\bar{u}_2$ which is thus a new constant of motion of the system.
We are then left with only one degree of freedom ($u_1$, $\bar{u}_1$) and the problem is integrable.

However, if one introduces second order terms in the Hamiltonian of a first order MMR,
this simplification no longer occurs.
It does not occur either, in the case of higher order MMR,
even at the minimal degree of development of the Hamiltonian.
Moreover, the presence of chaotic motion proves that no reduction
to an integrable system is possible \citep[see][]{wisdom_canonical_1986}.

For those higher order MMR, we have to make some additional hypothesis and/or
to neglect some terms in the Hamiltonian in order to obtain an integrable system that still capture
most of the characteristics of the resonant motion.
Nevertheless, we can get some inspiration from the method described for first order resonances.

The main idea of this method is to perform a rotation of coordinates and introduce
some kinds of proper modes of the motion ($u_i$). This rotation is chosen such as the fixed point
corresponding to the center of libration of the resonance \modif{lies} in the direction of the first
mode (i.e. $u_1\neq 0$, $u_2=0$ at the libration center).
Thus, all the resonant dynamics is concentrated in the first mode.
The second mode only adds some circulation around the libration center.

The aim of our method for higher order resonance is to obtain a similar rotation
that concentrate the main characteristics of the resonant motion in one mode.
Of course, we will have to make some simplifications and the model will not be able
to reproduce the chaotic motion that is observed in the numerical simulations.

\subsection{Higher order resonances}
\label{sec:higher-order-reson}

Let us now consider a resonance $(p+q)$:$p$ of order $q > 1$ and the Hamiltonian~(\ref{eq:Hq}).
The position of the center of libration can be obtained by solving Hamilton's equations with
zero right hand member \citep[e.g.][]{delisle_dissipation_2012}:
\modif{
  \begin{equation}
    \dot{x}_i = \ImUnit \dpart{\mathcal{H}}{\bar{x}_i} = 0
  \end{equation}
}
Let us note $I_{i,ell}$, $\sigma_{i,ell}$ the position of this libration center.
We then introduce the angle $\phi$ such that:
\begin{equation}
  \label{eq:phi}
  \tan \phi = \sqrt{\frac{I_{2,ell}}{I_{1,ell}}}
\end{equation}
and the diagonal matrix $M_\sigma$ with $(M_\sigma)_{i,i} = \expo{\ImUnit \sigma_{i,ell}}$.
It can easily be checked that the canonical change of coordinates:
\begin{equation}
  \label{eq:u}
  x = M_\sigma R_\phi u
\end{equation}
puts the libration center in the direction of $u_1$ ($u_2=0$).
Let us introduce the action-angle coordinates $D_i,\theta_i$
such that: $u_i = \sqrt{D_i}\expo{\ImUnit \theta_i}$.
The new Hamiltonian has exactly the same form as the Hamiltonian~(\ref{eq:Hq}):
\begin{equation}
  \label{eq:Hqb}
  \mathcal{H} = \mathcal{K}(\mathcal{D})
  + \mathcal{S}'_q(D_i,\theta_2-\theta_1)
  + \sum_{k=0}^q R'_k (u_1^ku_2^{q-k} + \bar{u}_1^k\bar{u}_2^{q-k})
\end{equation}
where we still have $\mathcal{D} = D_1 + D_2 = u_1\bar{u}_1 + u_2\bar{u}_2$.
In the case of first order resonances, the angle $\theta_2$ does not appear anymore in the Hamiltonian,
but here we still have a dependency on $\theta_2$ in the secular and the resonant parts.
In order to go further we need a simplifying approximation.
\modif{Let us recall that by construction, at the libration center, we have $D_{2,ell}=0$ while $D_{1,ell}\neq0$.
  Close to this elliptic fixed point, $D_2$ is negligible compared to $D_1$.
  The Hamiltonian~(\ref{eq:Hqb}) can be expanded in power series of $\sqrt{D_2/D_1}$, which is a small parameter
  as long as the system is close to the center of libration.
  We approximate the Hamiltonian~(\ref{eq:Hqb}) with the zeroth order part of this expansion that does not depend
  on $u_2$ but only on $u_1$. This is equivalent to imposing $u_2=0$ ($D_2 = 0$).
  In this approximation, the Hamiltonian is much simpler and integrable:
  \begin{equation}
    \mathcal{H} = \mathcal{K}(\mathcal{D})
    + \mathcal{S}'_q(D_1)
    + R'_q (u_1^q + \bar{u}_1^q)
  \end{equation}
  Moreover, we have $\mathcal{D} = D_1$ and the secular part becomes simply a polynomial of
  degree $\lfloor q/2 \rfloor$ in $D_1 = u_1\bar{u}_1$.
  Note that imposing $D_2 = 0$ is a strong hypothesis.
  Indeed, this implies that $I_2/I_1\approx \Lambda_2/\Lambda_1(e_2/e_1)^2$
  stays locked at its value at the libration center
  ($I_{2,ell}/I_{1,ell} = \tan^2\phi$).
  Therefore, the eccentricities stay very close to their values at the libration center.
  This hypothesis is reasonable when the system is close to the libration center.
  When the system is close to the separatrix, the dynamics is more complex, but we keep this
  simplified model as a first approximation.
}

\subsection{Development of the Keplerian part}
\label{sec:devel-kepl-part}

Since we obtained an integrable Hamiltonian there is no need for a new simplification.
However, if we study the dynamics at low eccentricities, the Keplerian part can be approximated with a polynomial in
$\mathcal{D}$ by a Taylor expansion. This allows one to obtain polynomials equations of motion.

We can first rewrite equations (\ref{eq:La1}), (\ref{eq:La2}) in the form:
\begin{eqnarray}
  \label{eq:La1b}
  \Lambda_1 &=& \Lambda_{1,0} - \frac{p}{q} (\mathcal{D} - \Delta G)\\
  \label{eq:La2b}
  \Lambda_2 &=& \Lambda_{2,0} + \frac{p+q}{q} (\mathcal{D} - \Delta G)
\end{eqnarray}
with $\Delta G = G_0 - G$, and $G_0$, $\Lambda_{i,0}$ the values at the nominal resonance
and zero eccentricities \citep[see][]{delisle_dissipation_2012}.
For a resonant system, the semi-major axis ratio (or the period ratio) should be close to the
nominal resonant value.
Therefore, we should have $\Lambda_i \approx \Lambda_{i,0}$ and thus $\mathcal{D} - \Delta G \approx 0$.
$\Delta G$ can thus be interpreted as an estimate (or a generalization)
of the total angular momentum deficit $\mathcal{D}$ \citep{laskar_spacing_2000}.
It should however be observed that $\Delta G$ is a constant of motion while $\mathcal{D}$
is not conserved.
Since, in the resonant case, $\mathcal{D} - \Delta G$ is small, we develop the Keplerian part
in power series of $\mathcal{D} - \Delta G$.
The constant terms of this development can be discarded because they have no impact on the dynamics.
Moreover, it can be shown that due to the definition of $\Lambda_{i,0}$ there is no contribution at first order.
The development should then be done at least at order two:
\begin{equation}
  \label{eq:devkep}
  \mathcal{K} \approx \mathcal{K}_0 - \mathcal{K}_2 \left(\mathcal{D} - \Delta G\right)^2
\end{equation}
where $\mathcal{K}_0$ is a constant that will be ignored in the following and
\begin{equation}
  \mathcal{K}_2 =  \frac{3}{2} \left(\frac{p}{q}\right)^2 n_{1,0}
  \left(\frac{1}{\Lambda_{1,0}}+\frac{p+q}{p}\frac{1}{\Lambda_{2,0}}\right)
\end{equation}
$n_{1,0}$ is the Keplerian mean motion at the nominal resonance:
\begin{equation}
  n_{1,0} = \frac{\mu_1^2\beta_1^3}{\Lambda_{1,0}^3}
\end{equation}

Note that the degree of development of the Keplerian part should be consistent
with the development of the perturbative part.
For a resonance of order $q$, we developed the perturbative part at degree $q$,
thus the Keplerian part should also be developed at order $q$ in eccentricity
(order $\lfloor q/2\rfloor$ in $\mathcal{D} - \Delta G$).
For the sake of simplicity we only consider here resonances of order $q<6$ for which the
Keplerian part can be approximated with:
\begin{equation}
  \mathcal{K} \approx -\mathcal{K}_2 D_1
  \left(D_1 - 2 \Delta G \right)
\end{equation}
where we eliminated constant terms and used the hypothesis $\mathcal{D} = D_1$ ($D_2 = 0$).

\subsection{Final form of the Hamiltonian}
\label{sec:final-form-hamilt}
The only remaining step is to express explicitly the secular part.
Since we assumed $q < 6$, $\mathcal{S}'_q$ has the form:
\begin{equation}
  \mathcal{S}'_q = S'_1 D_1 + S'_2 D_1^2
\end{equation}
(where we dropped constant terms) and the Hamiltonian reads:
\begin{equation}
  \mathcal{H} =  D_1 \bigg(2(\mathcal{K}_2 \Delta G + S'_1/2)
  - (\mathcal{K}_2-S'_2) D_1\bigg)
  + R'_q (u_1^q + \bar{u}_1^q)
\end{equation}

Let us note:
\begin{equation}
  \label{eq:delta}
  \delta = \frac{\mathcal{K}_2\Delta G + S'_1/2}{\mathcal{K}_2-S'_2} \approx \Delta G
\end{equation}
such that:
\begin{equation}
  \mathcal{H} =  (\mathcal{K}_2-S'_2) D_1 \left(2 \delta - D_1\right)
  + R'_q (u_1^q + \bar{u}_1^q)
\end{equation}

One can also rescale the Hamiltonian (and the timescale of the dynamics) with the transformation:
\begin{eqnarray}
  \mathcal{H}^* &=& \frac{\mathcal{H}}{\mathcal{K}_2-S'_2}\\
  R^* &=& \frac{R'_q}{\mathcal{K}_2-S'_2}\\
  t^* &=& (\mathcal{K}_2-S'_2) t
\end{eqnarray}

The final form we obtain for the Hamiltonian of a MMR of order q is:
\begin{equation}
  \label{eq:Hqc}
  \mathcal{H}^* = u_1\bar{u}_1 (2 \delta - u_1\bar{u}_1) + R^* (u_1^q + \bar{u}_1^q)
\end{equation}

In the following, we will drop the stars for the sake of brevity and will study the Hamiltonian:
\begin{equation}
  \label{eq:Hqd}
  \mathcal{H} = u_1\bar{u}_1 (2 \delta - u_1\bar{u}_1) + R (u_1^q + \bar{u}_1^q)
\end{equation}
or equivalently:
\begin{equation}
  \label{eq:Hqe}
  \mathcal{H} = D_1 (2 \delta - D_1) + 2 R D_1^{q/2} \cos(q\theta_1)
\end{equation}

For the sake of simplicity we will assume that $R > 0$.
If $R < 0$, one only needs to perform the rotation $\theta_1' = \theta_1 + \pi/q$ to meet
this hypothesis.

It is important to note that the change of coordinates $x \rightarrow u$ may depend on the parameter
$\Delta G$ since the direction of the libration center can change with $\Delta G$.
However, for second order resonances and at leading order in eccentricities, the direction always stays
the same and the change of coordinates is constant.
Moreover, for any MMR a constant change of coordinates can always be used as a first approximation
at low eccentricities
\citep[e.g][which gives the position of this libration center for most common MMR and a variety of mass ratio]{beauge_planetary_2006,michtchenko_stationary_2006}.
We thus suppose in the following that the change of coordinate is constant.

\subsection{Dynamics in the integrable model}
\label{sec:dynam-integr-model}

Using Hamilton equations and the Hamiltonian (\ref{eq:Hqd}) we obtain:
\begin{equation}
  \dot{u}_1 = \ImUnit \left( 2 u_1 (\delta - u_1\bar{u}_1) + q R \bar{u}_1^{q-1} \right)
\end{equation}
For $q>1$ there is an obvious fixed point at $u_1=0$, whereas for $q=1$ (first order MMR)
this fixed point never exists.

For $u_1\neq 0$, the position of the fixed points are given by:
\begin{equation}
  2 u_1\bar{u}_1(\delta - u_1\bar{u}_1) + q R \bar{u}_1^{q} = 0
\end{equation}
The first part of this equation is obviously real, thus $\bar{u}_1^q$ is also real.
This means that $\theta_1 = k \pi/q$ and $\bar{u}_1^q = (-1)^k |u_1|^q = (-1)^k D_1^{q/2}$.
We thus have to find the positive roots of:
\begin{equation}
  \label{eq:pfD1}
  \delta - D_1 + (-1)^{k} \frac{q}{2} R D_1^{q/2-1} = 0
\end{equation}

If $\delta$ is big enough, we can write $D_1 \sim \delta$.
At first order in $R$ the solution would be:
\begin{equation}
  \label{eq:pfD1b}
  D_1 = \delta + (-1)^{k} \frac{q}{2} R \delta^{q/2-1}
\end{equation}

Finally, the position of the fixed point in terms of $u_1$ can be approximated with:
\begin{equation}
  \label{eq:pfu1}
  u_1 = \sqrt{\delta + (-1)^k \frac{q}{2} R \delta^{q/2-1}} \expo{\ImUnit \frac{k \pi}{q}}
\end{equation}

\figI
\figII
Figure \ref{fig:I} shows the bifurcation diagram of fixed points for a second order resonance
with $R=0.1$ and Fig.~\ref{fig:II} shows the three different cases for the phase space depending
on the number of fixed points (1, 3, or 5).
For $\delta < -R$ (Fig.~\ref{fig:II}~top),
the system admits only one elliptical (stable) fixed point at zero (zero eccentricities)
and the phase-space exhibits only (external) circulation around this fixed point.
This clearly corresponds to a non-resonant (or secular) dynamics.
For $-R<\delta<R$ (Fig.~\ref{fig:II}~middle),
the fixed point at zero becomes hyperbolic (unstable) and
two symmetrical elliptical fixed points bifurcate from it on the real line.
Each one is associated to a resonant area.
The circulation area still exists, surrounding both resonant areas.
The separatrix between both types of motion passes through the hyperbolic point at zero.
Finally, for $\delta>R$ (Fig.~\ref{fig:II}~bottom),
the central fixed points becomes again stable and two symmetrical hyperbolic fixed points appear
on the imaginary line.
A new circulation area appears (internal circulation), between both resonant area, around the stable
fixed point at zero.

Higher order MMR \modif{exhibit} very similar behavior: successive bifurcations from the fixed point at zero.
However, the case of first order MMR is different.
Indeed, the fixed point at zero eccentricities does not exist for $q=1$, and the bifurcation
between purely secular motion and a resonant phase space significantly differs
\citep[see][]{henrard_second_1983,delisle_dissipation_2012}.
In particular, at the bifurcation, the center of external circulation becomes
the center of libration of the resonance.
This means that under dissipation (that induces an evolution of the phase space),
the system can evolve from external circulation to resonant motion (and vice versa)
without crossing the separatrix of the resonance.
This phenomenon was invoked to explain the excess of systems in \textit{Kepler} data that rely
close to first order MMR (2:1 and 3:2) but in external circulation \citep[see][]{delisle_dissipation_2012}.
For higher order resonances, the same process does not occur because the size of the resonant areas tends to
zero when reaching the bifurcation (see Figs.~\ref{fig:I},~\ref{fig:II}).
Thus a system always needs to cross the separatrix to pass from resonant motion to circulation.

\section{Dissipation in a resonant system}
\label{sec:diss-reson-syst}

\subsection{Amplitude of libration}
\label{sec:amplitude-libration}

The aim of this section is to study the evolution of a system that is initially trapped in resonance
and undergoes some dissipation, such as the tidal effect,
that damps the eccentricities of the planets.
At the beginning of the process, the system is assumed to be in the resonant area of the phase space
(with $\delta^2 \gg R\delta^{q/2}$, see Fig.~\ref{fig:II}~bottom).
We will assume here that the mode $u_1$ (defined in Eq.~(\ref{eq:u})) is damped on a timescale $T_d$:
\begin{equation}
  \label{eq:dud}
  \dot{u}_1|_d = - \frac{u_1}{T_d}
\end{equation}
and that this dissipative force induces a decrease of the parameter $\delta$
which is in first approximation proportional to the action $u_1\bar{u}_1$:
\begin{equation}
  \label{eq:ddd}
  \dot{\delta}|_d = -\gamma \frac{u_1\bar{u}_1}{T_d}
\end{equation}
We present in section~\ref{sec:outc-tidal-diss} estimates of $T_d$ and $\gamma$
in the case of tidal effect in both planets.

Let us introduce a measure of the amplitude of libration
in the resonant area:
\begin{equation}
  \label{eq:defA}
  A = \sin^2 \left(\frac{q\theta_{1,max}}{2}\right)
\end{equation}
where $\theta_{1,max}$ is the maximum value reached by the resonant angle during a libration.
We have $A=0$ at the center of libration, $A=1$ at the separatrix and $0<A<1$ in-between.
The evolution of the amplitude $A$ is governed by the following proposition:
\begin{proposition}
  \label{prop:I}
  Assuming that the conservative evolution of $u_1$ is given by the Hamiltonian~(\ref{eq:Hqd}),
  that the dissipation affects the system as described by Eqs.~(\ref{eq:dud}),(\ref{eq:ddd}),
  and that the system is in the regime $\delta^2 \gg R\delta^{q/2}$,
  the amplitude of libration in the resonance ($A$, Eq.~(\ref{eq:defA}))
  follows (demonstration in appendix~\ref{sec:evol-ampl}) :
  \begin{equation}
    \label{eq:dAdt}
    \dot{A}|_d \approx \frac{A}{T_d} \left( \left(1+\frac{q}{4}\right) \gamma - 2 \right)
  \end{equation}
\end{proposition}

Therefore, if $\gamma < 8/(4+q)$ the amplitude of libration decreases with respect to the size of the resonance,
while if $\gamma > 8/(4+q)$ the amplitude increases.
Let us note $\gamma_c$ this critical value:
\begin{equation}
  \label{eq:gammac}
  \gamma_c = \frac{8}{4+q}
\end{equation}

In the first case ($\gamma<\gamma_c$), the system evolves closer and closer to the fixed point.
On the long term, the system will follow very closely the stable branch while eccentricities tend
to zero.
On the other hand, if $\gamma>\gamma_c$, then $A$ grows exponentially and,
depending on the initial amplitude, the system can cross the separatrix before eccentricities are
completely damped.

After having crossed the separatrix, the system is no longer locked in resonance
and the semi-major axis ratio (or the period ratio) evolution depends on
the relative strength of the dissipation in each planet.

\subsection{Outcome of tidal dissipation in resonance}
\label{sec:outc-tidal-diss}

From our previous computations, it is clear that the most important parameter to determine
is $\gamma$ and especially how it compares with $\gamma_c$.
$T_d$ is also important since it gives the timescale
of the dissipative evolution and must be compared to the age of the considered system.

$\gamma$ is mainly influenced by the balance of the dissipation in both planets.
The more the first planet dissipates (compared to the second one) the greater is
$\gamma$.

At leading order in eccentricities the tidal effect induces \citep[e.g.][]{correia_tidal_2011}:
\begin{eqnarray}
  \label{eq:ed}
  \left.\dot{e}_i\right|_d &\approx& - \frac{e_i}{T_i}\\
  \label{eq:xd}
  \left.\dot{x}_i\right|_d &\approx& - \frac{x_i}{T_i}\\
  \label{eq:ad}
  \left.\dot{a}_i\right|_d &\approx& -2 e_i^2 \frac{a_i}{T_i}
  \approx - 4 \frac{x_i \bar{x}_i}{\Lambda_i} \frac{a_i}{T_i}
\end{eqnarray}
where $T_i$ is the timescale of the tidal dissipation in planet $i$.
For instance, if we take a constant lag time model \citep{singer_origin_1968,mignard_evolution_1979},
this timescale is given by
\citep[e.g.][]{bonfils_harps_2013}:
\begin{equation}
  \label{eq:Tid}
  T_i = \frac{2}{21} \frac{\beta_i a_i^8}{\Delta t_i k_{2,i} \mathcal{G} m_0^2 R_i^5}
\end{equation}
where $\Delta t_i$, $k_{2,i}$, and $R_i$ are the time lag, the second Love number,
and the radius of planet $i$.
The evolution of $u_1$ can be deduced from Eq.~(\ref{eq:xd}):
\begin{equation}
  \label{eq:u1d}
  \left.\dot{u}_1\right|_d =
  - \left(\frac{\cos^2\phi}{T_1} + \frac{\sin^2\phi}{T_2} \right) u_1
\end{equation}

The dissipation timescale $T_d$ is thus:
\begin{equation}
  \label{eq:Td}
  T_d = \frac{T_1 T_2}{T_1 \sin^2\phi + T_2 \cos^2\phi}
\end{equation}

The impact of such a dissipation on the parameter $\Delta G$ (at leading order) is given by
\citep[see][]{delisle_dissipation_2012}:
\begin{equation}
  \label{eq:dGd}
  \left.\dot{\Delta G}\right|_d =
  -2 G \left( \frac{p+q}{p} \frac{x_1\bar{x}_1}{T_1}
    + \frac{x_2\bar{x}_2}{T_2} \right)
\end{equation}
From which we deduce:
\begin{equation}
  \label{eq:dGd2}
  \left.\dot{\delta}\right|_d \approx \left.\dot{\Delta G}\right|_d =
  - \left(\frac{1}{1+\tau\tan^2\phi} \gamma_1
    + \frac{\tau\tan^2\phi}{1+\tau\tan^2\phi}\gamma_2\right) \frac{u_1 \bar{u}_1}{T_d}
\end{equation}
with:
\begin{eqnarray}
  \label{eq:tau}
  \tau &=& \frac{T_1}{T_2}\\
  \label{eq:gamma1}
  \gamma_1 &=& 2\frac{p+q}{p} G\\
  \label{eq:gamma2}
  \gamma_2 &=& 2 G
\end{eqnarray}
where $G$ can be approximated with $G_0$.
We thus have:
\begin{equation}
  \gamma = \frac{1}{1+\tau\tan^2\phi} \gamma_1 + \frac{\tau\tan^2\phi}{1+\tau\tan^2\phi}\gamma_2
\end{equation}

Since $(p+q)/p>1$, $\gamma_1 > \gamma_2$.
Therefore, $\gamma$ decreases with $\tau$,
from $\gamma_1$ ($\tau=0$, dissipation in planet 1 dominates)
to $\gamma_2$ ($\tau = +\infty$, dissipation in planet 2 dominates).

More precisely, it can be shown that
$\gamma_1 \in \left[2, 2(p+q)/p\right]$
and $\gamma_2 \in \left[2p/(p+q), 2\right]$.
The exact values depend on the masses of both planets. The lower bounds
correspond to the case $m_1 \ll m_2$ and the upper bounds to $m_2 \ll m_1$.
We thus always have (see Eq.~(\ref{eq:gammac})):
\begin{equation}
  \label{eq:g1c}
  \gamma_1 > \gamma_c
\end{equation}
Whereas, $\gamma_2$ can either be greater or smaller than $\gamma_c$ depending on
the considered \modif{resonance} and the masses of the planets.

If $\gamma_2 > \gamma_c$, then $\gamma > \gamma_c$ regardless of the balance of dissipation
in both planets ($\tau$).

If $\gamma_2 < \gamma_c$, then there exists a critical value of $\tau$:
\begin{equation}
  \label{eq:tauc}
  \tau_c = \frac{\gamma_1 - \gamma_c}{\gamma_c - \gamma_2}\cot^2\phi
\end{equation}
corresponding to the critical value $\gamma_c$.
If $\tau>\tau_c$, then $\gamma < \gamma_c$, and the amplitude of oscillation
decreases, whereas if $\tau<\tau_c$ the amplitude increases.

In the case $\tau > \tau_c$, the crossing of the separatrix is possible,
and the evolution of the semi-major axis ratio, $\alpha = a_1/a_2$, just after the
crossing can be estimated from Eq.(\ref{eq:ad}):
\begin{equation}
  \label{eq:alphad}
  \left.\frac{\dot{\alpha}}{\alpha}\right|_d \approx
  4 \frac{u_1 \bar{u}_1}{T_d}
  \left( \frac{\tau\tan^2\phi}{1+\tau\tan^2\phi}\frac{1}{\Lambda_2}
    - \frac{1}{1+\tau\tan^2\phi}\frac{1}{\Lambda_1} \right)
\end{equation}

Let us introduce:
\begin{equation}
  \label{eq:taua}
  \tau_\alpha = \frac{\Lambda_2}{\Lambda_1}\cot^2\phi
\end{equation}

Then, if $\tau > \tau_\alpha$, the semi-major axis ratio decreases (diverging orbits)
whereas if $\tau < \tau_\alpha$ the semi-major axis ratio increases (converging orbits).

Those two criteria can be rewritten in a more explicit form.
Let us recall that $\phi$ gives the direction of the libration center of the resonance.
More precisely, we have:
\begin{equation}
  \cot^2\phi = \frac{x_1\bar{x}_1}{x_2\bar{x}_2}
  \approx \frac{\Lambda_1 e_1^2}{\Lambda_2 e_2^2}
  \approx \frac{m_1 e_1^2}{m_2 e_2^2}\sqrt{\alpha}
\end{equation}
where all quantities must be evaluated at the libration center.

Thus $\tau_c$ and $\tau_\alpha$ can be approximated with:
\begin{eqnarray}
  \label{eq:tauc2}
  \tau_c &\approx& L \left(\frac{e_1}{e_2}\right)^2
  \frac{4 + (p+q)(1+L)}{4 L - p(1+L)}\\
  \label{eq:taua2}
  \tau_\alpha &\approx& \left(\frac{e_1}{e_2}\right)^2
\end{eqnarray}
where
\begin{equation}
  L = \frac{\Lambda_1}{\Lambda_2}
  \approx \frac{m_1}{m_2} \sqrt{\alpha}
  \approx \frac{m_1}{m_2} \left(\frac{p}{p+q}\right)^{1/3}
\end{equation}

To sum up:
\begin{itemize}
\item if $\tau\ (=T_1/T_2) > \tau_c$,
  the amplitude of libration decreases while the eccentricities are being damped.
  For first order resonances, the system leaves the resonance
  with diverging orbits when planets reach low eccentricities
  \citep[e.g.][]{delisle_dissipation_2012}.
  For higher order resonances the system stay in the resonant configuration.
\item if $\tau < \tau_c$,
  the amplitude of libration increases and the system may cross the separatrix depending
  on the initial amplitude.
  The smaller is $\tau$, the quicker the amplitude grows.
  If the initial amplitude is small and the growth is slow, the eccentricities are being damped
  before the system reaches the separatrix and it stays in resonance.
  Thus $\tau_c$ is a\modif{n} upper estimate of the maximal value of $\tau$ needed to cross the separatrix.
  The true maximal value depends on the initial eccentricities of the planets and on the initial
  amplitude of libration.
  If the system crosses the separatrix, the subsequent evolution
  is given by the comparison of $\tau$ and $\tau_\alpha$:
  \begin{itemize}
  \item if $\tau > \tau_\alpha$, the orbits are converging ($P_2/P_1<(p+q)/p$),
  \item if $\tau < \tau_\alpha$, the orbits are diverging ($P_2/P_1>(p+q)/p$).
  \end{itemize}
\end{itemize}

It should be noted that both $\tau_c$ and $\tau_\alpha$ have very simple expressions
(Eqs.~(\ref{eq:tauc2}),~(\ref{eq:taua2})) that do not depend on the
coefficient $R$ of the resonant part of the Hamiltonian.

The only difficulty in estimating $\tau_c$ and $\tau_\alpha$ is to estimate the eccentricity
ratio $e_1/e_2$ at the libration center.
This ratio (or equivalently the angle $\phi$) can be estimated from
the development of the Hamiltonian at leading degree but a better estimate can be obtained by
searching the fixed points of the Hamiltonian developed at a higher degree in eccentricities
\citep[e.g.][]{delisle_dissipation_2012},
or using numerical averaging methods
\citep[e.g.][]{beauge_planetary_2006,michtchenko_stationary_2006},
or computation\modif{s} of periodic orbits
\citep[e.g][]{hadjidemetriou_resonant_2002,antoniadou_resonant_2013}.

\section{Application to observed planetary systems}
\label{sec:applications}

In this section we present applications of our model to observed planetary systems.
In most planetary systems, the tidal dissipation is dominated by the contribution
of the inner planet, while the tidal effect in the outer planet can be neglected.
Indeed, the dissipation timescale has a strong dependency on the distance to the star (see Eq.~(\ref{eq:Tid})).
According to our model, in this case, the amplitude of libration always increases with time
(see Eq.~(\ref{eq:g1c})).
Therefore, if the initial amplitude is large enough, these systems can leave the resonance
by crossing the separatrix before eccentricities are damped.
Moreover, after these systems leave the resonance, the dissipation in the inner planet
induce an increase of the period ratio ($P_2/P_1$) since the semi-major axis of the inner
planet decreases (see Eq.~(\ref{eq:ad})).
We thus conclude that for most resonant planetary systems, the final outcome of the tidal dissipation
process should be external circulation if the initial amplitude of
libration was large enough, or resonant motion if the initial amplitude was small.

However, a system can end-up in internal circulation
if the tidal effect is much more efficient in the outer planet than in the inner one
(e.g. in the case of a gaseous inner planet and a rocky outer one), and/or if
the eccentricity of the outer planet is much larger than the inner planet's one.
Systems observed close to MMR, but in internal circulation are thus of particular interest
because we can obtain strong constraints on the nature of the planets (rocky or gaseous).
We looked in the exoplanet.eu database \citep{schneider_exoplanet_2011}
for systems observed in internal circulation
and with estimated masses compatible with a gaseous inner planet and a rocky outer one.
Only few systems correspond to these criteria.
We selected \object{GJ~163}b, c (3:1 MMR) and \object{GJ~581}b, c (5:2 MMR)
for illustrating our model.

\subsection{Application to GJ~163 b, c (3:1 MMR)}
\label{sec:appl-gj163}

\subsubsection{The detected system}
\label{sec:detected-system}

\object{GJ~163} is a M~dwarf that hosts 3 planets
\citep[see][and Table~\ref{tab:I} for the orbital parameters]{bonfils_harps_2013}.
The two inner ones are close to a 3:1 MMR with a period ratio of 2.97 (internal circulation).
The inner planet's minimum mass ($m\sin i$) is estimated to be 10.7 $M_\earth$,
while the second one is about 7.3 $M_\earth$.
The radii of these planets have not been estimated so
their density and nature (rocky or gaseous) is unknown.
However, it seems reasonable to suppose that the inner planet is gaseous while
the second one could be either rocky or gaseous \citep[see][]{bonfils_harps_2013}.
\tabI

\subsubsection{Scenario}
\label{sec:scenario}

The main question we want to answer is: ``Is there a natural explanation for having a system
very close but outside of the 3:1 MMR, with a period ratio $P_2/P_1 < 3$ (2.97) ?''.
It is of course possible that the system is close to the 3:1 MMR just by chance and that it
was never locked in this resonance.
However it seems more probable that the system has been locked in the 3:1 MMR
in the past (for instance due to convergent migration in the protoplanetary disk),
and afterwards slightly diverged from this resonant ratio to lower values.

We investigate here the possibility that tidal dissipation in planets may have induced this
resonant departure as described in sect.~\ref{sec:diss-reson-syst}.
Moreover we deduce constraints on the \object{GJ~163} system in order for this scenario to be valid.
\modif{Indeed, from sect.~\ref{sec:diss-reson-syst} we deduce that for the system to leave
  the resonance with $P_2/P_1<(p+q)/p$,
  the ratio $\tau = T_1/T_2$ of tidal dissipation timescales in both planets must verify:
  $\tau_\alpha < \tau < \tau_c$ (see Eqs.~(\ref{eq:taua2}) and~(\ref{eq:tauc2})
  for definitions of $\tau_\alpha$ and $\tau_c$).
  For numerical applications, we use a constant lag time model \citep{singer_origin_1968,mignard_evolution_1979}.
  Criteria on $\tau$ transcribe in criteria on the
  lag time ratio $\Delta t_2/\Delta t_1$ by using the expression of $T_i$ given in Eq.~(\ref{eq:Tid}):
  \begin{equation}
    \label{eq:rDtftau}
    \frac{\Delta t_2}{\Delta t_1} = \frac{1}{\alpha^8}\frac{\beta_2}{\beta_1}
    \frac{k_{2,1}}{k_{2,2}} \left(\frac{R_1}{R_2}\right)^5 \tau
    \approx \frac{1}{\alpha^8} \frac{m_2}{m_1} \kappa \tau
  \end{equation}
  with
  \begin{equation}
    \kappa = \frac{k_{2,1}}{k_{2,2}} \left(\frac{R_1}{R_2}\right)^5
  \end{equation}
  and
  \begin{equation}
    \alpha = \frac{a_1}{a_2}
  \end{equation}
}

For this application, the eccentricity ratio is computed using the simplest analytical model
(degree 2).
We obtain $e_1/e_2 \approx 1.1$ at the libration center.
\modif{
  The criterion $\tau > \tau_\alpha$ implies $\Delta t_2/\kappa\Delta t_1\gtrsim 300$.
  The criterion $\tau < \tau_c$ implies $\Delta t_2/\kappa\Delta t_1\lesssim 1450$.
  Thus the scenario that we described should be possible if:
  \begin{equation}
    \label{eq:Dtrange}
    300 \lesssim \frac{\Delta t_2}{\kappa\Delta t_1} \lesssim 1450
  \end{equation}
}
\subsubsection{N-body Simulations}
\label{sec:n-body-simulations}

\figIII

We performed numerical simulations of \object{GJ~163}, starting in resonance
with different lag time ratios for the tidal effect in both planets.
We used a constant $\Delta t$ model for the dissipation \citep{singer_origin_1968,mignard_evolution_1979}
and the ODEX integrator \citep[e.g.][]{hairer_solving_2010} as described in \citet{bonfils_harps_2013}.
Gaseous planets typically have dissipation quality factors $Q \sim 10^3 - 10^4$,
while rocky planets have quality factors in the range $Q \sim 10 - 100$.
This corresponds to lag times in the range $\Delta t \sim 10 - 100$ s for gaseous planets,
and $\Delta t \sim 10^3 - 10^4$ s for rocky ones ($Q \approx 1/(n \Delta t)$, where $n$ is the
mean-motion of the planet).
Since the innermost planet is probably gaseous, its lag time should be in the range
$\Delta t_1 \sim 10 - 100$ s.
However, in order to speed up the simulations, we used a higher tidal lag time.
As explained by \citet{bonfils_harps_2013}, the timescale of the evolution is roughly inversely
proportional to $\Delta t$.
In order to check that this approximation does not affect dramatically our results we performed
simulations with $\Delta t_1 = 10^7$, $10^6$, $10^5$, and $10^4$ s
(respectively 5, 4, 3, and 2 orders of magnitude higher than the expected value).
The system is integrated during respectively 0.1, 1, 10, and 100 Myr which would roughly correspond
to 10 Gyr with $\Delta t_1 = 100$ s.
As a comparison, the age of the system is estimated to be in the range $1-10$ Gyr
\citep{bonfils_harps_2013}.

\figIV

We chose the initial elliptical elements of the planets such that the system is initially in
resonance but not exactly at the libration center.
Indeed, as explained in sect.~\ref{sec:diss-reson-syst}, when the initial amplitude of libration
is small, the system may not cross the separatrix before eccentricities are very small
even if the amplitude increases.
This initial amplitude is generated by varying the mean anomaly of the inner planet
while all other elliptical elements correspond to an ACR (libration center of the resonance).
For $M_1=0^\circ$, the system is initially at the libration center while for $M_1=180^\circ$,
it starts at the separatrix (see Fig.\ref{fig:III} for a diagram of initial conditions).
The initial eccentricities of the planets are set to about 0.16 and 0.11 in order to be close to the ACR,
the perihelia are anti-aligned and $M_2=0^\circ$.
We took $a_1=0.062$ such that when eccentricities goes to zero the system end-up approximately at
its current position ($a_1\approx 0.0607$).

\figV

Fig.~\ref{fig:IV} shows the final period ratio of the planets as a function of
the lag time ratio with $M_1 = 100^\circ$ (top) and $140^\circ$ (bottom).
In both cases, we superimposed the results obtained with the four different dissipative timescales
($\Delta t_1 = 10^7$, $10^6$, $10^5$, and $10^4$ s).
We see in Fig.~\ref{fig:IV} that the four curves show very similar features
and exhibit only small variations between them.
We do not observe a particular trend when varying the timescale of the dissipation.
Thus, we can assume that taking a dissipation timescale several order of magnitudes
higher than realistic values does not affect much the results of this study,
while it speeds up the computations.

As described by our model, we observe in Fig.~\ref{fig:IV} the three possible final states
(external/internal circulation or resonant motion) for the
system depending on the value of \modif{$\Delta t_2/\kappa\Delta t_1$}.
We plot in Fig.~\ref{fig:V} an example of simulation
for each of the three final configurations.
For each case, we plot the evolution of the period ratio, the eccentricities, and the resonant angles.
The limit between external and internal circulation (Fig.~\ref{fig:IV}) is around
\modif{$\Delta t_2/\kappa\Delta t_1 \approx 250$} for both initial amplitudes of libration
($M_1 = 100$ and $140^\circ$).
Our analytical estimate (300, see Eq.~(\ref{eq:Dtrange})) is fairly close to this numerical computation.
The limit between internal circulation and resonant motion depends (as expected)
on the initial amplitude and occurs at \modif{$\Delta t_2/\kappa\Delta t_1 \approx 750$} for $M_1=100^\circ$,
and 900 for $M_1=140^\circ$.
The analytical model value (1450, see Eq.~(\ref{eq:Dtrange}))
is thus significantly higher than these numerical results.
However, as explained in sect.~\ref{sec:diss-reson-syst} and observed in the simulation,
this limit highly depends on the initial amplitude and/or initial eccentricities.
In our simple model we do not take into account these parameters and obtain an upper estimate
of this limit which corresponds to a system initially in resonance but close to the separatrix
($M_1\rightarrow 180^\circ$).

In order to illustrate this dependency we performed numerical simulations of the system with
different initial amplitude of libration ($M_1$).
The final period ratio of the system as a function of the lag time ratio and the initial amplitude
is shown in Fig.\ref{fig:VI}.
For this figure we used $\Delta t_1 = 10^7$ s since we showed that this value does not affect
strongly our results and it allows to speed-up \modif{the} computations.

For a small initial amplitude of libration, we observe that the system always end-up in resonance
(green zone in Fig.~\ref{fig:VI}).
For higher initial amplitudes, internal (in blue) and external (in yellow/orange) circulation are possible.
We observe that the limit between external and internal circulation always occurs for
\modif{$\Delta t_2/\kappa\Delta t_1 \approx 250$}.

The limit between internal circulation and resonance depends on
the initial amplitude of libration.
At very high initial amplitudes, the limit tends to
\modif{$\Delta t_2/\kappa\Delta t_1 \sim 1100$}.
This is \modif{fairly close} to the analytically predicted value (1450, see Eq.~(\ref{eq:Dtrange})).
\modif{The remaining difference between analytical and numerical results probably comes from the
  simplifying hypothesis we used in the model.
  Indeed, analytical results are obtained by assuming that $e_1/e_2$ stays locked
  at its value at the libration center.
  While this is reasonable when the system is close to the libration center,
  when it is close to the separatrix,
  the eccentricities can significantly differ from their values at the libration center.}

It should be noted that around $M_1 = 120^\circ$, some systems,
which were expected to stay in resonance,
end-up in internal circulation (blue points in the green area).
This is due to the presence of the separatrix of the secular resonance inside
the mean-motion resonance.
Such resonances were already studied in the case of the 2:1 and 3:2 MMR by
\citet{callegari_dynamics_2004,callegari_dynamics_2006}.
As described by these studies, the secular frequency (dominating the motion of $\Delta \varpi$)
falls to zero near the separatrix and the angle $\Delta \varpi$
librates in the opposite direction from one side to the other of the separatrix.
Of course, our \modif{simplified} model cannot predict this kind of phenomenon.
However, \modif{it} still captures the main
features of the resonant motion under dissipation and allows a better understanding of the mechanisms
that lead to the three different final states in our simulations.

It is important to notice that what we call initial amplitude of libration is the initial
condition of our specific simulations.
If we had chosen lower initial eccentricities, the amplitude of libration
would have less time to grow, and a larger initial amplitude would be necessary to cross the
separatrix of the resonance before reaching zero eccentricities.
With higher initial eccentricities, the minimal initial amplitude of libration
needed to cross the separatrix would be smaller.
Moreover, at moderate eccentricities ($\sim 0.3$),
the phase space of the resonance exhibits bifurcations \citep[e.g.][]{michtchenko_stationary_2006}
and when the system crosses these bifurcations the amplitude of libration undergoes jumps.
This means that an even smaller initial amplitude of libration would be necessary if the system
was initially at larger eccentricities.

\modif{
  If the system was initially in a 3:1 resonance,
  our model implies that $\Delta t_2/\Delta t_1 \sim 500 \kappa$.
  In order to conclude on the nature of both planets, we need to estimate $\kappa = k_{2,1}/k_{2,2} (R_1/R_2)^5$.
  The Love numbers of both planets should be of the same order of magnitude\footnote{
    \modif{In the Solar System, the Love numbers ($k_2$) of the Earth is 0.29 \citep{kozai_love_1968}, Jupiter 0.379, Saturn 0.341 \citep{gavrilov_love_1977}}}.
  The radius ratio, raised to the fifth power, has a more significant impact.
  Since the planets radii are unknown we estimate them from the masses using the empirical power law
  obtained by \citet{weiss_mass_2013}:
  \begin{equation}
    \frac{R_1}{R_2} \approx \left(\frac{m_1}{m_2}\right)^{0.53} \left(\frac{a_1}{a_2}\right)^{0.06}
  \end{equation}
  Using this estimate for \object{GJ~163}b, c, we obtain :
  \begin{equation}
    \label{eq:rhogj163}
    \kappa \approx \left(\frac{R_1}{R_2}\right)^5 \approx 2.2
  \end{equation}
  The upper x-axis of Fig.~\ref{fig:VI} is scaled using this value of $\kappa$.
  Finally, we conclude that $\Delta t_2/\Delta t_1 \sim 1000$.
  Such a ratio is reasonable only if the first planet is gaseous while the second one is telluric
  \citep[see][]{bonfils_harps_2013}.
  It should be noted that the minimum mass of the outer planet ($m\sin i$) is about 7 $M_\earth$.
  If it is a rocky planet, its mass is probably close to this value, and its inclination should be close to $90^\circ$.
}

\figVI

\subsection{Application to GJ~581 b, c (5:2 MMR)}
\label{sec:gj581}

The model and mechanisms we describe in this study are very general and are not limited
to the \object{GJ~163} planetary system.
Our conclusions are valid for a MMR of any order $q$.
However, the criterion we obtain on the lag time ratio (in both planets) depends
on the considered MMR and on the masses of the planets which
influence the position of the libration center of the resonance.
Thus, the final outcome of the dissipation (internal/external circulation or resonant motion)
must be determined for each system individually.

In order to illustrate the generality of our mechanism,
we performed a similar analysis on the system \object{GJ~581}.
This system has raised much discussion about the number of detected planets.
Most studies agree on the presence of 4 planets around this M-dwarf and both planets b and c,
which we are interested in, are uncontested.
We reproduced the orbital parameters of the 4 planets system given in \citet{forveille_only_2011} in Table~\ref{tab:I}.
Planets b, and c \citep[see][]{bonfils_neptune_2005,udry_super_2007,mayor_earth_2009,forveille_only_2011}
have a period ratio of 2.4 which is close to a 5:2 MMR (internal circulation).
The inner planet (b) has a minimum mass of about $15.9 M_\earth$ while planet c
has a minimum mass of $5.3 M_\earth$.
Applying our analytical criterion to this system we deduce that internal circulation is
possible for \modif{$\Delta t_2/\kappa\Delta t_1 \in [3,52]$} (with $e_1/e_2 \approx 0.25$).
As for \object{GJ~163}, we performed numerical simulations of the system for different
lag time ratios \modif{($\Delta t_2/\kappa\Delta t_1$)} and different initial amplitude of libration ($M_1$).
We set the initial conditions to be close to the center of libration of the resonance.
The eccentricities of the planets are initially set to about 0.05 and 0.2, the perihelia are
anti-aligned, and $M_2 = -36^\circ$.
We took $a_1 = 0.042$ such that when eccentricities are damped the system end-up approximately at its current position
($a_1\approx 0.041$).
For $M_1=0^\circ$, the system is initially at the center of libration,
while for $M_1=90^\circ$ the system begins on the separatrix.
The system is integrated during $5\times 10^4$~yr with $\Delta t_1 = 10^7$~s.
The age of the system is estimated to be about 8 Gyr, thus our simulations
are approximately equivalent to $\Delta t_1 \sim 10-100$ s
on the age of the system (which corresponds to a gaseous planet).

The final outcome of these simulations are represented in Fig.~\ref{fig:VII}.
We observe a good agreement between analytical estimates and numerical results.
In both case, the current configuration of \object{GJ~581}b and c is obtained if
\modif{$\Delta t_2/\Delta t_1 \sim 20\kappa$}.
It is interesting to note that this lag time ratio is significantly (one order of magnitude) smaller than
what we obtained for \object{GJ~163}.
The main reason for this is the differences in the position of the libration center in terms of eccentricity.
For \object{GJ~163}, we have $e_1/e_2 \approx 1.1$ while here we have $e_1/e_2 \approx 0.25$ at the libration center.
Since the lag time ratio is proportional to the eccentricity ratio squared this explain the difference between both results.
It should be observed that for most resonances, at low eccentricities, $e_1/e_2$ increases with $m_2/m_1$
\citep[e.g.][]{michtchenko_stationary_2006}.

\modif{As for \object{GJ~163}, the radii and Love numbers of both planets are unknown.
  We suppose equal Love numbers for both planets and estimate the radii
  using the same empirical power law as for \object{GJ~163} \citep{weiss_mass_2013}:
  \begin{equation}
    \label{eq:rhogj581}
    \kappa \approx \left(\frac{R_1}{R_2}\right)^5 \approx 15
  \end{equation}
  The top x-axis of Fig.~\ref{fig:VII} is scaled using this value of $\kappa$.
  The lag time ratio of this system should be $\Delta t_2/\Delta t_1 \sim 300$.
  As for \object{GJ~163}, we conclude that, if \object{GJ~581}b, c formed by tidal dissipation
  in the 5:2 MMR, the inner planet should be gaseous and the outer one rocky.
  The minimum masses (respectively $15.86\ M_\earth$ and $5.34\ M_\earth$
  for the inner and the outer planets) are compatible with this conclusion.
}
\figVII

\section{\textit{Kepler}'s statistics}
\label{sec:kepler}

Multi-planetary systems detected by the \textit{Kepler} mission
present an excess of planet pairs close to first order MMR
(2:1 and 3:2), but in external circulation
\citep[see][]{lissauer_architecture_2011,fabrycky_architecture_2012}.
Since tidal dissipation in planets involved in a first order MMR can induce a departure
from the resonance to external circulation when eccentricities reach very low values,
this scenario has been proposed to explain \textit{Kepler}'s statistics
\citep{lithwick_resonant_2012,delisle_dissipation_2012,batygin_dissipative_2013}.
However, the timescale of this dissipation might be too long compared to the age of the systems to explain
their present configurations (see \citealp{lee_kepler_2013} and also \citealp{rein_period_2012}).
It should be noted that the scenario considered in these studies assumes that the planets
leave the resonance with very low eccentricities and without crossing the separatrix of the resonance.
The departure of the period ratio from the resonant value is very slow
since the tidal dissipation decreases with eccentricities (see Eqs.~(\ref{eq:ed})-(\ref{eq:ad})).

Nevertheless, a resonant system that initially has a large enough amplitude of libration can
cross the separatrix of the resonance while the eccentricities of the planets are still high.
The departure of the period ratio from the resonant value is much faster in this case since the eccentricities are higher.
As we observed in the introduction of sect.~\ref{sec:applications}, in most planetary systems
the tidal dissipation mainly occurs in the inner planet and the dissipation
in the outer one can be neglected.
In this case, the amplitude of libration increases with time, and when a system crosses
the resonance separatrix, the period ratio increases.
Therefore, this mechanism also produces the excess of planets in external circulation observed in
\textit{Kepler} data, but on a shorter timescale.

Neglecting the dissipation in the outer planet, the evolution of the period ratio
($\mathcal{P} = P_2/P_1$) after the system crossed the separatrix is given by:
\begin{equation}
  \frac{\dot{\mathcal{P}}}{\mathcal{P}} = 3 e_1^2 \frac{1}{T_1}
\end{equation}

For the sake of simplicity we will neglect the secular interactions between the planets and
suppose that $e_1$ undergoes an exponential decrease $e_1 = e_{1,0} \expo{-t/T_1}$, with $e_{1,0}$
the eccentricity of the inner planet when the system crosses the separatrix,
and $T_1$ the dissipation timescale (see Eq.~(\ref{eq:Tid})).
In this approximation, the temporal evolution of the period ratio follows:
\begin{equation}
  \label{eq:PP0}
  \log(\mathcal{P}/\mathcal{P}_0) = \frac{3}{2} e_{1,0}^2 \left(1 - \expo{-2t/T_1}\right)
\end{equation}
with $\mathcal{P}_0 = (p+1)/p$.

Using Eq.~(\ref{eq:PP0}), one can obtain the maximum value reached by the period ratio:
\begin{equation}
  \label{eq:Plim}
  \mathcal{P} = \frac{P_2}{P_1}\ _{\overrightarrow{t \to \infty}}\ \mathcal{P}_0 \expo{3/2 e_{1,0}^2}
\end{equation}
Figure~\ref{fig:VIII} shows this maximum value as a function of the eccentricity $e_{1,0}$.
It should be noted that when eccentricities reach low values the mechanism invoked previously
\citep[e.g.][]{delisle_dissipation_2012} dominates
and the period ratio continues to increase but much slowly.

\figVIII

\citet{lee_kepler_2013} estimated that tidal dissipation would need at least $t \gtrsim 50 T_1$
for the system to reach $\mathcal{P} - \mathcal{P}_0 \approx 0.03$,
both for the 2:1 and the 3:2 resonances.
Depending on the considered resonance and on the masses of both planets,
this can even reach $t \gtrsim 1000\ T_1$.

\figIX

Using Eq.~(\ref{eq:PP0}) we can estimate the time needed for the system to reach
the same configuration but supposing it left the resonance with significant eccentricities.
We plot in Fig.~\ref{fig:IX} this estimate ($t/T_1$) as a function of the eccentricity $e_{1,0}$.
We see in Fig.~\ref{fig:IX} a vertical asymptote at $e_{1,0} \approx 0.1$.
This is because, for $e_{1,0} \lesssim 0.1$,
the limit $\lim\limits_{t\to\infty}\mathcal{P}-\mathcal{P}_0$ is smaller than 0.03.
In this case, the system can eventually reach 0.03, on a timescale shorter but comparable to
the one estimated by \citet{lee_kepler_2013} (50-1000 $T_1$).
On the \modif{opposite}, when $e_{1,0} \gtrsim 0.15$, the system reaches the desired
period ratio on a much shorter timescale: $t/T_1 \sim 0.01 - 0.1$.
This correspond to a gain of 3-5 orders of magnitude with this \modif{alternate} scenario.

\citet{lee_kepler_2013} discarded many near-resonant systems because tidal dissipation
seemed too slow to explain their current configuration.
However, this study only considered a scenario of resonant departure at low eccentricities.
Some of these discarded systems might actually have formed by crossing
the resonance separatrix with non-negligible eccentricities ($e_{1,0} \gtrsim 0.15$)
due to the increase of the amplitude of libration induced by the tidal dissipation.
In that case, the evolution of the period ratio after the resonance breaking is about 3-5 orders of magnitude
more rapid, and the current configuration can be obtained on a more reasonable timescale.

\section{Conclusion}
\label{sec:conclusion}

We presented an integrable model of mean-motion resonances of any order.
This model is highly simplified and cannot reproduce all the features of the
resonant dynamics.
However it allows to deduce a very simple criterion on the tidal dissipation undergone by
both planets to end-up inside the resonance, or on a side or the other of the resonance.
The main factors that enter into account are the balance of tidal dissipation between both planets
($T_1/T_2$ or $\Delta t_2/\Delta t_1$) and the position of the libration center
(especially the ratio $e_1/e_2$).

\modif{Using this criterion on the two inner planets orbiting \object{GJ~163} we deduce that the current
  period ratio (2.97) could be the outcome of dissipation in the 3:1 MMR provided that
  $\Delta t_2/\Delta t_1 \sim 1000$.}
Using N-body simulations with dissipation we reach the same conclusion with slightly refined
bounds for $\Delta t_2/\Delta t_1$.
Both methods clearly imply that the inner planet
should be gaseous and the outer planet should be rocky.
The minimum masses of both planets (respectively 10.7 $M_\earth$ and 7.3 $M_\earth$) are compatible with
this hypothesis, but since the inclinations and the radii are currently unknown,
some uncertainty remains.
\modif{We also applied this model to \object{GJ~581}b, c and could reproduce the current configuration
  with tidal dissipation in the 5:2 MMR if $\Delta t_2/\Delta t_1 \sim 300$.
  As for \object{GJ~163}, we conclude that the inner planet
  should be gaseous and the outer planet should be rocky, which is compatible with
  the minimum masses of both planets (respectively 15.86 $M_\earth$ and 5.34 $M_\earth$
  for the inner and the outer planets).
}

As we noted in the case of \object{GJ~163}, some secondary resonances can affect the outcome of the
considered system.
\modif{Our integrable model of resonances is not able to predict such a complex behavior,
  as well as chaotic motion.
  This might be a limitation for high order resonances, which may show large chaotic areas.}
Moreover, we make all our estimates using a constant eccentricity ratio ($e_1/e_2$) which
is computed at the center of libration of the resonance.
As eccentricities are being damped, the position of the libration center evolves
and the eccentricity ratio is not constant.
Depending on the resonance and on the considered range of eccentricities,
the changes on the eccentricity ratio at the libration center can be non-negligible
\citep[e.g.][]{michtchenko_stationary_2006}.
Besides, when the amplitude of libration is small, eccentricities of both planets should be
close to the values at the libration center, but when the system reaches the separatrix and
leaves the resonance, the eccentricities can be significantly different.
Therefore, this estimate of the eccentricity ratio is the main limit in our model
and in the computation of criteria on the lag time ratio.
However, as we observed in the cases of \object{GJ~163} and \object{GJ~581},
with this approximation we still obtain a good estimate
of the order of magnitude of the lag time ratio
and a better understanding of the mechanisms
that are at stake in determining the outcome of the dissipative process.

The most interesting cases to study are those in internal circulation because
we can obtain strong constraints on the nature of the planets for our scenario
to be possible (as for \object{GJ~163} and \object{GJ~581}).
However, our mechanism also applies to many systems that are observed in
external circulation.
Indeed, in most cases, the tidal dissipation in the outer planet is negligible
compared to the dissipation in the inner planet, and the most probable outcome for the system
is external circulation.
For first order MMR, external circulation can also be obtained when eccentricities are very low
without crossing the separatrix of the resonance
\citep[the separatrix simply disappear\modif{s} at low eccentricities, e.g.][]{delisle_dissipation_2012}.
However, the subsequent evolution of the period ratio is very slow due to the smallness of the eccentricities.
\citet{lee_kepler_2013} showed that for many systems the evolution of the period ratio is too slow to
reach the current value on a reasonable timescale.
We \modif{show} that a gain of 3-5 orders of magnitude on the timescale is obtained by considering a scenario
of resonance breaking due to tides at non-negligible eccentricities ($e_1 \gtrsim 0.15$).
\modif{This allows to explain the presence of an excess of planets
  in external circulation in \textit{Kepler} data without
  introducing any other mechanism than tidal dissipation.}

\begin{acknowledgements}
  \modif{We thank the anonymous referee for pointing us the importance of planet radii in the estimate of the tidal dissipation
    and other constructive comments that improved the quality of this article.}
  This work has been supported by PNP-CNRS,
  CS of Paris Observatory,
  PICS05998 France-Portugal program,
  and FCT-Portugal (PEst-C/CTM/LA0025/2011).
\end{acknowledgements}

\bibliographystyle{aa}
\bibliography{DLC}

\begin{thebibliography}{38}
\expandafter\ifx\csname natexlab\endcsname\relax\def\natexlab#1{#1}\fi

\bibitem[{{Antoniadou} \& {Voyatzis}(2013)}]{antoniadou_resonant_2013}
{Antoniadou}, K.~I. \& {Voyatzis}, G. 2013, \apss

\bibitem[{{Batygin} \&
  {Morbidelli}(2013{\natexlab{a}})}]{batygin_analytical_2013}
{Batygin}, K. \& {Morbidelli}, A. 2013{\natexlab{a}}, \aap, 556, A28

\bibitem[{{Batygin} \&
  {Morbidelli}(2013{\natexlab{b}})}]{batygin_dissipative_2013}
{Batygin}, K. \& {Morbidelli}, A. 2013{\natexlab{b}}, \aj, 145, 1

\bibitem[{Beaug\'{e} {et~al.}(2006)Beaug\'{e}, Michtchenko, \&
  {Ferraz-Mello}}]{beauge_planetary_2006}
Beaug\'{e}, C., Michtchenko, T.~A., \& {Ferraz-Mello}, S. 2006, Monthly Notices
  of the Royal Astronomical Society, 365, 1160

\bibitem[{{Bonfils} {et~al.}(2005){Bonfils}, {Forveille}, {Delfosse}, {Udry},
  {Mayor}, {Perrier}, {Bouchy}, {Pepe}, {Queloz}, \&
  {Bertaux}}]{bonfils_neptune_2005}
{Bonfils}, X., {Forveille}, T., {Delfosse}, X., {et~al.} 2005, \aap, 443, L15

\bibitem[{{Bonfils} {et~al.}(2013){Bonfils}, {Lo Curto}, {Correia}, {Laskar},
  {Udry}, {Delfosse}, {Forveille}, {Astudillo-Defru}, {Benz}, {Bouchy},
  {Gillon}, {H\'{e}brard}, {Lovis}, {Mayor}, {Moutou}, {Naef}, {Neves}, {Pepe},
  {Perrier}, {Queloz}, {Santos}, \& {S\'{e}gransan}}]{bonfils_harps_2013}
{Bonfils}, X., {Lo Curto}, G., {Correia}, A.~C.~M., {et~al.} 2013, \aap, 556,
  A110

\bibitem[{Callegari {et~al.}(2006)Callegari, {Ferraz-Mello}, \&
  Michtchenko}]{callegari_dynamics_2006}
Callegari, N., {Ferraz-Mello}, S., \& Michtchenko, T.~A. 2006, Celestial
  Mechanics and Dynamical Astronomy, 94, 381

\bibitem[{Callegari {et~al.}(2004)Callegari, Michtchenko, \&
  {Ferraz-Mello}}]{callegari_dynamics_2004}
Callegari, N., Michtchenko, T.~A., \& {Ferraz-Mello}, S. 2004, Celestial
  Mechanics and Dynamical Astronomy, 89, 201

\bibitem[{{Correia} {et~al.}(2011){Correia}, {Laskar}, {Farago}, \&
  {Bou\'{e}}}]{correia_tidal_2011}
{Correia}, A.~C.~M., {Laskar}, J., {Farago}, F., \& {Bou\'{e}}, G. 2011,
  Celestial Mechanics and Dynamical Astronomy, 111, 105

\bibitem[{{Delisle} {et~al.}(2012){Delisle}, {Laskar}, {Correia}, \&
  {Bou\'{e}}}]{delisle_dissipation_2012}
{Delisle}, J.-B., {Laskar}, J., {Correia}, A.~C.~M., \& {Bou\'{e}}, G. 2012,
  \aap, 546, A71

\bibitem[{{Fabrycky} {et~al.}(2012){Fabrycky}, {Lissauer}, {Ragozzine}, {Rowe},
  {Agol}, {Barclay}, {Batalha}, {Borucki}, {Ciardi}, {Ford}, {Geary}, {Holman},
  {Jenkins}, {Li}, {Morehead}, {Shporer}, {Smith}, {Steffen}, \&
  {Still}}]{fabrycky_architecture_2012}
{Fabrycky}, D.~C., {Lissauer}, J.~J., {Ragozzine}, D., {et~al.} 2012, ArXiv
  e-prints

\bibitem[{{Forveille} {et~al.}(2011){Forveille}, {Bonfils}, {Delfosse},
  {Alonso}, {Udry}, {Bouchy}, {Gillon}, {Lovis}, {Neves}, {Mayor}, {Pepe},
  {Queloz}, {Santos}, {Segransan}, {Almenara}, {Deeg}, \&
  {Rabus}}]{forveille_only_2011}
{Forveille}, T., {Bonfils}, X., {Delfosse}, X., {et~al.} 2011, ArXiv e-prints

\bibitem[{{Gavrilov} \& {Zharkov}(1977)}]{gavrilov_love_1977}
{Gavrilov}, S.~V. \& {Zharkov}, V.~N. 1977, \icarus, 32, 443–449

\bibitem[{{Goldreich} \& {Schlichting}(2014)}]{goldreich_overstable_2014}
{Goldreich}, P. \& {Schlichting}, H.~E. 2014, \aj, 147, 32

\bibitem[{{Hadjidemetriou}(2002)}]{hadjidemetriou_resonant_2002}
{Hadjidemetriou}, J.~D. 2002, Celestial Mechanics and Dynamical Astronomy, 83,
  141

\bibitem[{Hairer {et~al.}(2010)Hairer, N{\o}rsett, \&
  Wanner}]{hairer_solving_2010}
Hairer, E., N{\o}rsett, S.~P., \& Wanner, G. 2010, {Solving Ordinary
  Differential Equations I: Nonstiff Problems} (Springer)

\bibitem[{{Henrard} \& {Lemaitre}(1983)}]{henrard_second_1983}
{Henrard}, J. \& {Lemaitre}, A. 1983, Celestial Mechanics, 30, 197

\bibitem[{{Henrard} {et~al.}(1986){Henrard}, {Milani}, {Murray}, \&
  {Lemaitre}}]{henrard_reducing_1986}
{Henrard}, J., {Milani}, A., {Murray}, C.~D., \& {Lemaitre}, A. 1986, Celestial
  Mechanics, 38, 335

\bibitem[{{Kozai}(1968)}]{kozai_love_1968}
{Kozai}, Y. 1968, \pasj, 20, 24

\bibitem[{{Laskar}(2000)}]{laskar_spacing_2000}
{Laskar}, J. 2000, Physical Review Letters, 84, 3240

\bibitem[{Laskar \& Robutel(1995)}]{laskar_stability_1995}
Laskar, J. \& Robutel, P. 1995, Celestial Mechanics and Dynamical Astronomy,
  62, 193

\bibitem[{{Lee} {et~al.}(2013){Lee}, {Fabrycky}, \& {Lin}}]{lee_kepler_2013}
{Lee}, M.~H., {Fabrycky}, D., \& {Lin}, D.~N.~C. 2013, \apj, 774, 52

\bibitem[{{Lissauer} {et~al.}(2011){Lissauer}, {Ragozzine}, {Fabrycky},
  {Steffen}, {Ford}, {Jenkins}, {Shporer}, {Holman}, {Rowe}, {Quintana},
  {Batalha}, {Borucki}, {Bryson}, {Caldwell}, {Carter}, {Ciardi}, {Dunham},
  {Fortney}, {Gautier}, {Howell}, {Koch}, {Latham}, {Marcy}, {Morehead}, \&
  {Sasselov}}]{lissauer_architecture_2011}
{Lissauer}, J.~J., {Ragozzine}, D., {Fabrycky}, D.~C., {et~al.} 2011, \apjs,
  197, 8

\bibitem[{{Lithwick} \& {Wu}(2012)}]{lithwick_resonant_2012}
{Lithwick}, Y. \& {Wu}, Y. 2012, \apjl, 756, L11

\bibitem[{{Mayor} {et~al.}(2009){Mayor}, {Bonfils}, {Forveille}, {Delfosse},
  {Udry}, {Bertaux}, {Beust}, {Bouchy}, {Lovis}, {Pepe}, {Perrier}, {Queloz},
  \& {Santos}}]{mayor_earth_2009}
{Mayor}, M., {Bonfils}, X., {Forveille}, T., {et~al.} 2009, \aap, 507, 487

\bibitem[{Michtchenko {et~al.}(2006)Michtchenko, Beaug\'{e}, \&
  {Ferraz-Mello}}]{michtchenko_stationary_2006}
Michtchenko, T.~A., Beaug\'{e}, C., \& {Ferraz-Mello}, S. 2006, Celestial
  Mechanics and Dynamical Astronomy, 94, 411

\bibitem[{{Michtchenko} \& {Ferraz-Mello}(2001)}]{michtchenko_modeling_2001}
{Michtchenko}, T.~A. \& {Ferraz-Mello}, S. 2001, \icarus, 149, 357–374

\bibitem[{{Mignard}(1979)}]{mignard_evolution_1979}
{Mignard}, F. 1979, Moon and Planets, 20, 301

\bibitem[{{Novak} {et~al.}(2003){Novak}, {Lai}, \&
  {Lin}}]{novak_interesting_2003}
{Novak}, G.~S., {Lai}, D., \& {Lin}, D.~N.~C. 2003, in {Astronomical Society of
  the Pacific Conference Series}, Vol. 294, {Scientific Frontiers in Research
  on Extrasolar Planets}, ed. D.~{Deming} \& S.~{Seager}, 177--180

\bibitem[{Papaloizou(2011)}]{papaloizou_tidal_2011}
Papaloizou, J. C.~B. 2011, Celestial Mechanics and Dynamical Astronomy, 111, 83

\bibitem[{Papaloizou \& Terquem(2010)}]{papaloizou_dynamics_2010}
Papaloizou, J. C.~B. \& Terquem, C. 2010, Monthly Notices of the Royal
  Astronomical Society, 405, 573

\bibitem[{{Rein}(2012)}]{rein_period_2012}
{Rein}, H. 2012, \mnras, 427, L21

\bibitem[{{Schneider} {et~al.}(2011){Schneider}, {Dedieu}, {Le Sidaner},
  {Savalle}, \& {Zolotukhin}}]{schneider_exoplanet_2011}
{Schneider}, J., {Dedieu}, C., {Le Sidaner}, P., {Savalle}, R., \&
  {Zolotukhin}, I. 2011, \aap, 532, A79

\bibitem[{{Sessin} \& {Ferraz-Mello}(1984)}]{sessin_motion_1984}
{Sessin}, W. \& {Ferraz-Mello}, S. 1984, Celestial Mechanics, 32, 307

\bibitem[{Singer(1968)}]{singer_origin_1968}
Singer, S.~F. 1968, Geophysical Journal of the Royal Astronomical Society, 15,
  205

\bibitem[{{Udry} {et~al.}(2007){Udry}, {Bonfils}, {Delfosse}, {Forveille},
  {Mayor}, {Perrier}, {Bouchy}, {Lovis}, {Pepe}, {Queloz}, \&
  {Bertaux}}]{udry_super_2007}
{Udry}, S., {Bonfils}, X., {Delfosse}, X., {et~al.} 2007, \aap, 469, L43

\bibitem[{{Weiss} {et~al.}(2013){Weiss}, {Marcy}, {Rowe}, {Howard}, {Isaacson},
  {Fortney}, {Miller}, {Demory}, {Fischer}, {Adams}, {Dupree}, {Howell},
  {Kolbl}, {Johnson}, {Horch}, {Everett}, {Fabrycky}, \&
  {Seager}}]{weiss_mass_2013}
{Weiss}, L.~M., {Marcy}, G.~W., {Rowe}, J.~F., {et~al.} 2013, \apj, 768, 14

\bibitem[{{Wisdom}(1986)}]{wisdom_canonical_1986}
{Wisdom}, J. 1986, Celestial Mechanics, 38, 175

\end{thebibliography}

\appendix

\section{Evolution of the amplitude of libration}
\label{sec:evol-ampl}

The aim of this appendix is to prove Prop.~\ref{prop:I}.
Assuming $\delta^2\gg R\delta^{q/2}$, the Hamiltonian~(\ref{eq:Hqe}) can be approximated with the
pendulum-like Hamiltonian:
\begin{equation}
  \label{eq:Hpendul}
  \mathcal{H} = D_1(2\delta - D_1) + 2 R \delta^{q/2} \cos(q\theta_1)
\end{equation}
Noting $D_1 = \delta + \epsilon$, we obtain:
\begin{equation}
  \label{eq:Hpendulb}
  \mathcal{H} = \delta^2 - \epsilon^2 + 2 R \delta^{q/2} \cos(q\theta_1)
\end{equation}
Hamilton's equations give (in the conservative case):
\begin{eqnarray}
  \label{eq:HamEqpend}
  \dot\theta_1 &=& -2 \epsilon\\
  \dot\epsilon &=& 2 q R \delta^{q/2} \sin(q\theta)
\end{eqnarray}
The elliptical fixed point corresponds to $q \theta_{1,ell} = 0$, $\epsilon_{ell} = 0$, $\mathcal{H}_{ell} = \delta^2 + 2 R \delta^{q/2}$.
The hyperbolic fixed point is at $q \theta_{1,hyp} = \pi$, $\epsilon_{hyp} = 0$, $\mathcal{H}_{hyp} = \delta^2 - 2 R \delta^{q/2}$.
Noting $\Delta \mathcal{H} = \mathcal{H}_{ell} - \mathcal{H}_{hyp} = 4R\delta^{q/2}$, one can verify that:
\begin{equation}
  \label{eq:Hpendulc}
  \mathcal{H}_{ell} - \mathcal{H} = \epsilon^2 +\Delta\mathcal{H}\sin^2\left(\frac{q\theta_1}{2}\right)
\end{equation}

For a given conservative resonant trajectory, the maximum value of $\theta_1$ is reached when $\dot\theta_1=-2\epsilon = 0$.
This corresponds to:
\begin{equation}
  \label{eq:thetamax}
  \sin^2\left(\frac{q\theta_{1,max}}{2}\right) = \frac{\mathcal{H}_{ell}-\mathcal{H}}{\Delta\mathcal{H}}
\end{equation}
This clearly provide a measure of the amplitude of libration which we call $A$:
\begin{equation}
  \label{eq:Ampli}
  A  = \sin^2\left(\frac{q\theta_{1,max}}{2}\right)
  = \frac{\mathcal{H}_{ell}-\mathcal{H}}{\Delta\mathcal{H}}
  = \frac{\epsilon^2}{\Delta \mathcal{H}} + \sin^2\left(\frac{q\theta_1}{2}\right)
\end{equation}

Let us now consider the evolution of this amplitude under dissipation.
The dissipation affects the system as described by Eqs.~(\ref{eq:dud}),(\ref{eq:ddd}).
From these expressions we derive:
\begin{eqnarray}
  \dot{D}_1|_d &=& - 2 \frac{D_1}{T_d} = - \frac{2}{T_d} (\delta + \epsilon)\\
  \dot\delta|_d &=& -\frac{\gamma}{T_d} (\delta + \epsilon)\\
  \dot\epsilon|_d &=& \dot{D}_1|_d - \dot\delta|_d = \frac{\gamma-2}{T_d}(\delta+\epsilon)\\
  \dot{\Delta \mathcal{H}}|_d &=& -\frac{q}{2}\frac{\gamma}{T_d}\Delta \mathcal{H}\left(1 + \frac{\epsilon}{\delta}\right)
\end{eqnarray}

The instantaneous derivative of the amplitude $A$ is given by (see Eq.~(\ref{eq:Ampli})):
\begin{eqnarray}
  \label{eq:dAdtb}
  \dot A|_d &=& \frac{\mathrm{d}}{\mathrm{d}t}
  \left.\left(\frac{\epsilon^2}{\Delta\mathcal{H}}\right)\right|_d\\
  &=& \frac{1}{\Delta\mathcal{H}^2} (2\Delta\mathcal{H}\epsilon\dot\epsilon|_d - \epsilon^2\dot{\Delta\mathcal{H}}|_d)\\
  \label{eq:dAdtc}
  &=& \frac{2}{T_d\Delta \mathcal{H}}\left(
    (\gamma-2)\delta\epsilon
    + \left(\left(1+\frac{q}{4}\right)\gamma-2\right)\epsilon^2
    + \gamma \frac{q}{4}\frac{\epsilon^3}{\delta}
  \right)
\end{eqnarray}

Let us average this instantaneous derivative over one libration period in order to
evaluate the long term evolution of $A$.
We need to compute the mean values of $\epsilon$, $\epsilon^2$, and $\epsilon^3$.
Due to the symmetry of the problem, $\epsilon$ and $\epsilon^3$ average out to zero
(odd powers of $\epsilon$).
We thus simply have:
\begin{equation}
  \label{eq:dAdtd}
  <\dot A|_d> = \frac{2<\epsilon^2>}{T_d\Delta \mathcal{H}}\left(
    \left(1+\frac{q}{4}\right)\gamma-2\right)
\end{equation}
with
\begin{equation}
  \label{eq:eps2}
  <\epsilon^2> = \left(\int_0^{\theta_{1,max}} \frac{\epsilon^2}{|\dot\theta_1|}\mathrm{d}\theta_1\right)/
  \left(\int_0^{\theta_{1,max}} \frac{1}{|\dot\theta_1|}\mathrm{d}\theta_1\right)
\end{equation}
and
\begin{eqnarray}
  \dot\theta_1 &=& -2\epsilon\\
  \epsilon^2 &=& \Delta\mathcal{H}\left(\sin^2\left(\frac{q\theta_{1,max}}{2}\right)
    -\sin^2\left(\frac{q\theta_1}{2}\right)\right)
\end{eqnarray}
Both integrals can be expressed using elliptic integrals and we obtain:
\begin{eqnarray}
  \label{eq:eps2}
  <\epsilon^2> &=& \Delta \mathcal{H}\left( A + \frac{E\left(\sqrt{A}\right)}{K\left(\sqrt{A}\right)} - 1\right)\\
  &\approx& \frac{\Delta \mathcal{H} A}{2}
\end{eqnarray}
where $K$ and $E$ are the complete elliptic integrals of the first and second kinds.
Finally, at leading order, the evolution of the amplitude $A$ is governed by:
\begin{equation}
  <\dot{A}|_d> \approx \frac{A}{T_d} \left( \left(1+\frac{q}{4}\right) \gamma - 2 \right)
\end{equation}
\end{document}